\RequirePackage{lineno}    
\documentclass[10pt]{article}
\usepackage{amsmath}
\usepackage{amssymb}
\usepackage{graphicx}
\usepackage{cite}
\usepackage{color} 
\usepackage{booktabs}
\usepackage{ctable}

\usepackage{setspace} 
\usepackage[labelfont=bf,labelsep=period,justification=raggedright]{caption}
\makeatletter
\renewcommand{\@biblabel}[1]{\quad#1.}
\makeatother

\begin{document}
\begin{flushleft} 
{\Large \textbf{A model of autophagy size selectivity by receptor clustering on peroxisomes}}\\
Aidan I Brown$^{1,2}$,  
Andrew D Rutenberg$^{1,\ast}$\\
$^1$Department of Physics and Atmospheric Science, Dalhousie University, Halifax, Nova Scotia, Canada B3H 4R2\\
$^2$Department of Physics, Simon Fraser University, Burnaby, British Columbia, Canada, V5A 1S6\\
$\ast$ E-mail: andrew.rutenberg@dal.ca
\end{flushleft}
\today

\section*{Abstract}%
Selective autophagy must not only select the correct type of organelle, but also must discriminate between individual organelles of the same kind so that some but not all of the organelles are removed. We propose that physical clustering of autophagy receptor proteins on the organelle surface can provide an appropriate all-or-none signal for organelle degradation.  We explore this proposal using a computational model restricted to peroxisomes and the relatively well characterized pexophagy receptor proteins NBR1 and p62. We find that larger peroxisomes nucleate NBR1 clusters first and lose them last through competitive coarsening. This results in significant size-selectivity that favors large peroxisomes, and can explain the increased catalase signal that results from siRNA inhibition of p62.  Excess ubiquitin, resulting from damaged organelles, suppresses size-selectivity but not cluster formation. Our proposed selectivity mechanism thus allows all damaged organelles to be degraded, while otherwise selecting only a portion of organelles for degradation.

\newpage
\section{Introduction}
Macroautophagy (hereafter autophagy) can degrade large subcellular substrates such as organelles and pathogenic bacteria \cite{stolz14, rogov14, randow14}. Selective autophagy can target specific damaged or surplus substrates \cite{green14}, and defects in this process often lead to human disease \cite{mizumura14}.  To degrade targeted substrates, the downstream stages of autophagy include formation of and recruitment to phagophores, and formation and maturation of the autophagosome.  The initial targeting of substrates involves an ``eat-me'' signal such as ubiquitin and subsequent recruitment of autophagy receptor proteins, with only some receptor types recruited for a given type of organelle \cite{rogov14}. 

Selective autophagy involves an all-or-none response, where each substrate is either targeted or not for degradation, leading to a relatively very high or relatively low respective rate of degradation by the autophagy system. Such all-or-none responses can be produced through cooperative effects \cite{gutierrez12, kholodenko06}. Clustering is a cooperative mechanism that can generate a qualitative all-or-none response, as illustrated by the regulation of bacteriophage lysis timing where lysis follows only after the collective formation of holin-clusters on the cell surface \cite{ryan07, white11}.  The involvement of protein clusters in autophagy is supported by reports of domains of distinct receptor proteins on bacteria targeted for xenophagy \cite{cemma11, wild11, mostowy11}.  

Autophagy receptor proteins are relatively well characterized for mammalian peroxisomes. Peroxisomes are essential and dynamic cellular organelles with functions including  metabolism of hydrogen peroxide and  oxidation of fatty acids \cite{islinger12}. Peroxisomes are approximately spherical with diameters ranging from $\sim 0.1 - 1\mu$m \cite{smith13}. There can be hundreds of peroxisomes in a single mammalian cell \cite{ezaki11}. In both mammals and yeast significant peroxisome degradation is through autophagy, known as pexophagy \cite{iwata06, monastryska04}. In mammalian autophagy, peroxisomes are targeted for degradation by exogenous ubiquitin labeling \cite{kim08}. However, the receptor protein NBR1 is also both necessary and sufficient for pexophagy \cite{deosaran13} while the receptor p62 significantly contributes though is not essential \cite{kim08, deosaran13}.  The J region of NBR1 \cite{deosaran13} allows NBR1 to anchor directly to organelle membranes. 

We propose that a sufficient all-or-none signal for autophagy selectivity can be provided by the initial formation of receptor clusters on substrate surfaces.  Specifically, for mammalian peroxisomes, we hypothesize that NBR1 can form lateral clusters once it is associated with membranes, and that an NBR1 cluster on a peroxisome is necessary for downstream degradation while the absence of an NBR1 cluster prevents downstream degradation. In this paper we explore this cluster-selectivity hypothesis in the context of mammalian peroxisomes. Small clusters of receptor proteins, if initially placed on every subcellular organelle, will subsequently grow and shrink due to receptor exchange between organelles \cite{brown15}.  With such initial cluster placement, selectivity would only emerge at late times when only a few organelles are left with clusters of receptors.  We consider here the selective initial formation of NBR1 clusters, the role of ubiquitin on NBR1 recruitment, and the possible effect of p62 on NBR1 cluster formation. We take a computational modeling approach to investigating our hypotheses, so that we can quantitatively test the self-consistency of our model \cite{chakraborty10}.  

Our computational model is deliberately simple with respect to the complex biological regulation of cellular processes.  Our goal is simply to demonstrate a physically viable mechanism by which autophagy selectivity might operate, consistent with the known biophysical functions of the peroxisomal receptor proteins NBR1 and p62. With this computational approach, we both pose and address three essential questions about autophagy selectivity: (1) what mechanism can provide an all-or-none signal to target individual organelles for degradation by autophagy, (2) how can this mechanism respond flexibly to organelle damage, and (3) what needs to be measured in the lab to better characterize mechanisms of autophagy selectivity? We also identify robust qualitative implications of our hypotheses that can be confronted with experiments, even in the face of considerable parameter uncertainty \cite{gutenkunst07, papin05}. 
    
\section{Results}
\subsection{NBR1 model description}

\begin{figure}[h] 
	\centering
	\hspace{-0.3in}
		\hspace{-0.0in} \includegraphics[width=6.0in]{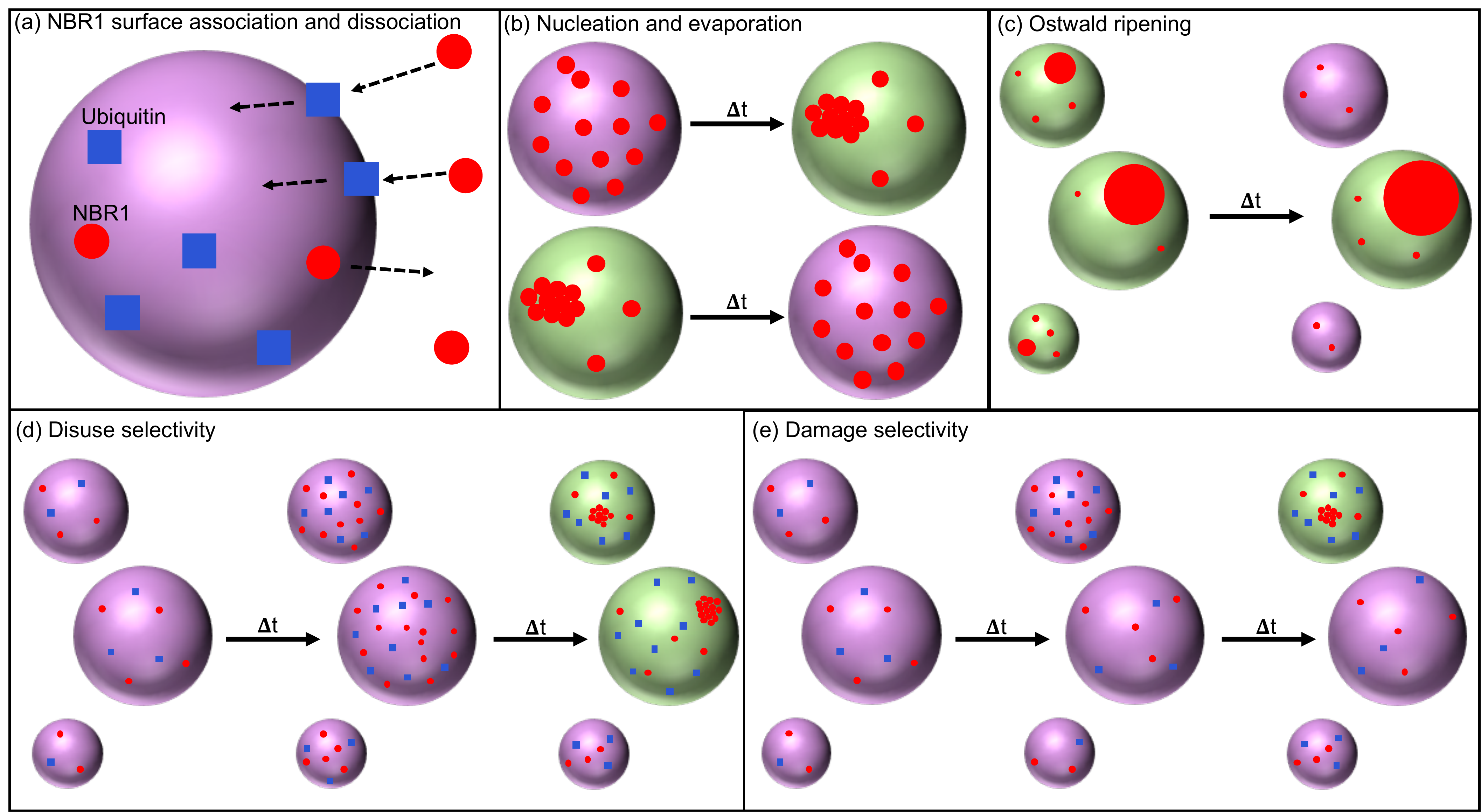} 
	\caption{\label{fig:diagram}  {\bf Model processes:}
(a) NBR1 (red circles) in the cytosol surrounding a peroxisome (large purple sphere) associates with the peroxisome membrane via transient association with ubiquitin (blue squares) bound to the peroxisome membrane (association rate in Eqn.~\ref{eq:jonnbr1}). NBR1 on the peroxisome membrane can then dissociate from the peroxisome membrane (dissociation rate in Eqn.~\ref{eq:steadystateNBR1}).
(b) With sufficient NBR1 present on the peroxisome membrane, a cluster of NBR1 nucleates (required concentration in Eqn.~\ref{eq:thresholdconcentration}). The number of NBR1 in a cluster can increase or decrease following nucleation (Eqn.~\ref{eq:dynamicalmain}). Peroxisomes coloured green harbour an NBR1 cluster and may be selected for degradation by autophagy.
(c) Ostwald ripening results in the growth of larger clusters, and shrinking and evaporation of smaller clusters (Eqn.~\ref{eq:dynamicalmain} describes cluster size dynamics).
(d) Under ``disuse'' conditions that cause the number of ubiquitin on all peroxisomes to increase, more NBR1 will associate with peroxisome membrane, facilitating cluster nucleation, and the selection of larger peroxisomes for degradation by autophagy.
(e) Damage may cause the number of ubiquitin on a subset of peroxisomes to increase, leading to increased NBR1 association on those peroxisomes,  possible cluster nucleation, and selection of those damaged peroxisomes for degradation by autophagy.}
\end{figure}
 
We start with an NBR1-only model, since NBR1 is the only peroxisomal receptor that is both necessary and sufficient for mammalian pexophagy \cite{deosaran13}. The results of our NBR1 model development are presented here without mathematical details (see details below in Computational Methods).  Our model for NBR1 dynamics has three aspects: NBR1 association with the peroxisome surface, formation of  NBR1 clusters from freely-diffusing  NBR1 on the peroxisome surface, and the growth or shrinkage of  NBR1 clusters once they have formed. These aspects are illustrated in Fig.~\ref{fig:diagram} (a)-(c).

Peroxisomes have surface-displayed ubiquitin \cite{brown14, kim08}, and NBR1 have a ubiquitin binding region (UBA) \cite{lin13, kraft10, vadlamudi96}, so NBR1 can associate with peroxisomes by attaching to the surface-displayed ubiquitin. However, the dissociation constant of the UBA region is much larger than estimated NBR1 concentrations --- strongly  suggesting that NBR1 will only transiently associate with ubiquitin (see details in Computational Methods). This leaves the amphipathic J domain of NBR1 \cite{deosaran13} as the  dominant mode of lasting association, following much lower dissociation constants for amphipathic helices.  

We model NBR1 recruitment to the peroxisome surface as diffusion-limited arrival to and transient association with the ubiquitin on the peroxisome surface via the UBA domain of NBR1, immediately followed by long-lived association with the peroxisome membrane via the J region of NBR1. This is quantitatively described by a standard equation for diffusion-limited association to absorbing targets on a sphere, see Eqn.~\ref{eq:jonnbr1} below.

Once associated with the peroxisomal membrane, our model allows NBR1 to self-associate into homo-oligomeric clusters, motivated by the coiled-coil domains of NBR1 \cite{deosaran13, kirkin09}, the importance of NBR1 to aggregation formation \cite{nicot14}, and generic aggregation phenomenon driven by non-specific interactions  \cite{heimburg96, gil98, lague01, bray02}. Formation of  clusters from many freely-diffusing individual NBR1 already on the peroxisome surface occurs at a critical concentration of individual NBR1. To determine this critical concentration for cluster formation we require sufficient NBR1 to both form a cluster (condition 1) and leave behind enough individual NBR1 to prevent the cluster from immediately evaporating (condition 2). The critical concentration is the lowest concentration at which both of these conditions can be satisfied, given by Eqn.~\ref{eq:minimumc} below, which results in the formation of a cluster.

Our model of NBR1 cluster size increase and decrease is in line with the standard physical picture for such processes, known as Ostwald ripening \cite{yao93, voorhees85}. Clusters described by Ostwald ripening will shrink if they are below some threshold size, and grow if they are above the threshold size, with this threshold size growing in time \cite{yao93, voorhees85}. The continuous evolution of an individual cluster size with time $t$ is given by
\begin{equation}
	\label{eq:dynamicalmain}
	\frac{dN_{clust}}{dt} = 4\pi a R^2\left[w - f_{\infty}\left(1 + \nu\sqrt{\frac{\pi}{bN_{clust}}}\right)\right],
\end{equation}
 which is derived below leading up to Eqn.~\ref{eq:dynamical}.  $N_{clust}$ is the number of NBR1 in a cluster, and $R$ is the radius of the peroxisome harbouring the cluster. The total cellular NBR1 which is not associated with peroxisomes is in a shared cellular pool, and it is through this pool that NBR1 can exchange between different peroxisomes.  Two variables, $a(t)$ and $w(t)$, are time-dependent combinations of variables such as the cellular concentration of NBR1, or the number of freely-diffusing NBR1 on the peroxisome under consideration. The remaining parameters  ($f_\infty$, $\nu$, and $b$) are constants, independent of time or the peroxisome under consideration. 

\subsection{NBR1 Results}

\begin{figure}[tbp] 
	\centering
	\hspace{-0.3in}
	\begin{tabular}{c}
		\hspace{-0.00in}\includegraphics[width=3.0in]{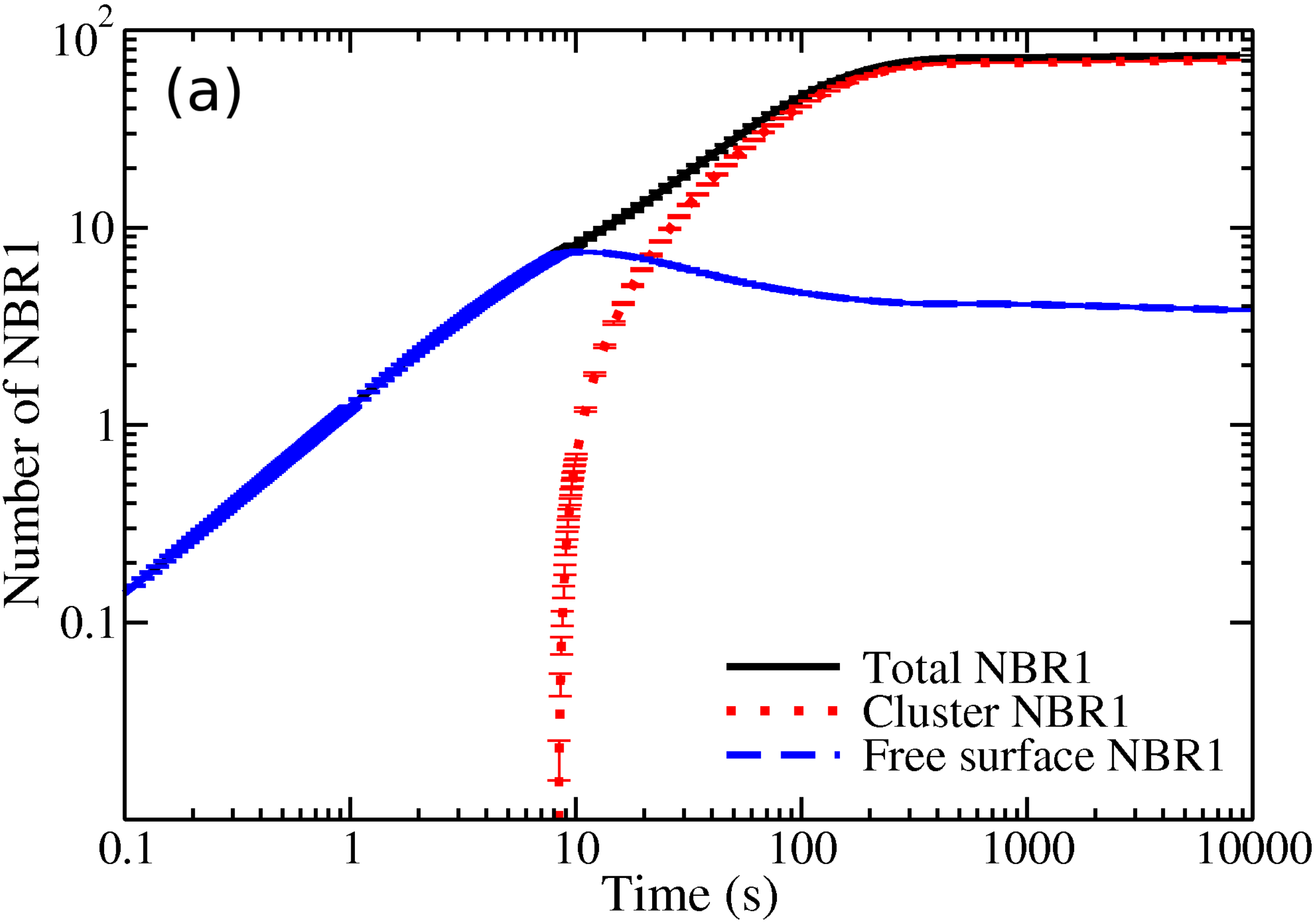} \\
		\hspace{-0.00in}\includegraphics[width=3.0in]{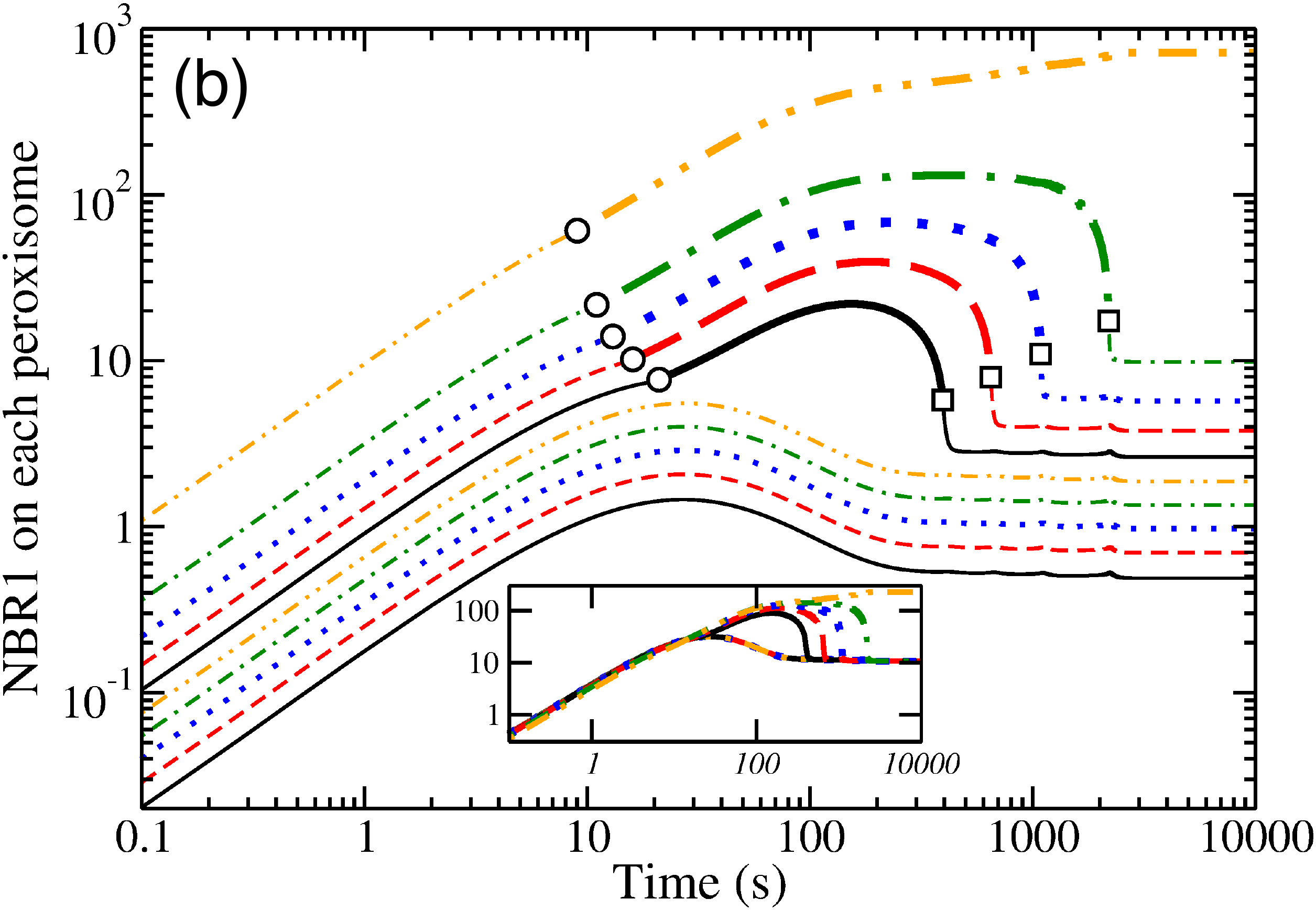}\\
		\hspace{-0.0in} \includegraphics[width=3.0in]{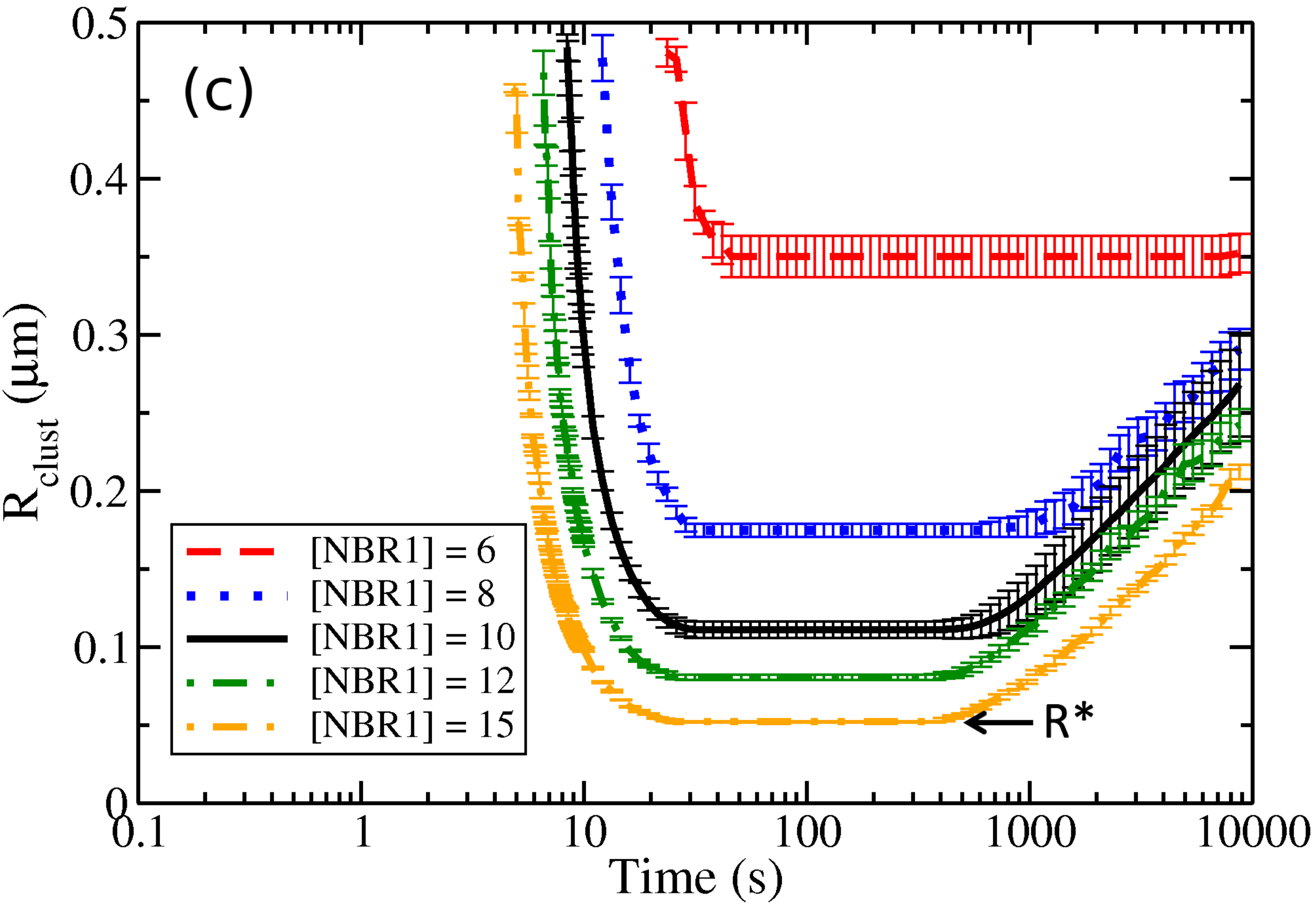} \\
	\end{tabular}
	\caption{\label{fig:figure1}  (a) The amount of NBR1 associated with a peroxisome of radius $R=0.25$ $\mu$m vs.\ time. The initial ($t=0$) bulk concentration is [NBR1] = 10 $\mu$m$^{-3}$, with no surface NBR1.  At approximately time $t=10$ s, a cluster is nucleated and grows thereafter.  Cluster growth (dotted red line) allows the total NBR1 (solid black line) to greatly exceed the saturation level exhibited by the free NBR1 (dashed blue line).  (b) The total amount of NBR1 vs.\ time is shown for each of ten peroxisomes, with radii exponentially distributed between $R=0.05$ $\mu$m and $R=0.50$ $\mu$m.  The largest peroxisomes have the most NBR1, and the smallest the least NBR1. Each curve is narrow when no cluster is present on each peroxisome, and thick when a cluster is present. Cluster nucleation is indicated by circles, evaporation by squares. Only the largest five peroxisomes form clusters, starting around $t=20$ s. For all but the largest peroxisome, the clusters nucleate, grow, and then shrink again as the clusters coarsen \cite{brown15}. Inset of (b) shows the NBR1 surface concentration on peroxisomes vs.\ time ---  the concentrations for the different peroxisomes, all of different radii, are equal except when a cluster is present.  (c) The size of the smallest peroxisome with a cluster, $R_{clust}$, vs.\ time shows how at early and late times only the largest peroxisomes have clusters. At intermediate times, clusters form and remain on all but the smallest peroxisomes. $R_{clust}$ reaches a minimum at intermediate times, indicating the radius of the smallest peroxisome to be occupied by a cluster at any time, $R^*$. The legend indicates the initial bulk NBR1 concentration -- higher concentrations lead to lower $R^*$.}
\end{figure}

Using our quantitative NBR1 model, we examine NBR1 dynamics on individual peroxisomes. This includes the behavior of the number of associated NBR1 with time, the timing of the formation of NBR1 clusters at a critical concentration, and the growth of clusters after they have formed. 

\subsubsection{NBR1 clusters form promptly and allow much more NBR1 to associate with peroxisomes} 
When peroxisomes are placed in a medium with a bulk concentration of NBR1 (at time $t=0$), Fig.~\ref{fig:figure1}(a) shows how the recruitment of NBR1 by  ubiquitin leads to an approximately linear increase of freely diffusing surface-associated NBR1 (dashed blue line). At very early times there are no NBR1 clusters (dotted red line), so the total NBR1 on the peroxisome (solid black line) consists only of NBR1 freely diffusing on the surface. However, once the surface NBR1 concentration surpasses the critical concentration of nucleation, at approximately $t=10$ s, a cluster is nucleated (top process in Fig.~\ref{fig:diagram}(b)) and very quickly grows in size. At late times almost the entire population of surface-associated NBR1 is in clusters, while the population of freely-diffusing NBR1 decreases somewhat after cluster nucleation. This decrease is due to the reduced supersaturation necessary to maintain a larger cluster as compared to the supersaturation necessary to nucleate a smaller cluster (see Eqn.~\ref{eq:formationsecond} below and \cite{krishnamachari96}).

The behavior of freely-diffusing and cluster-bound NBR1 illustrates how the existence of clusters can lead to a prompt all-or-none response --- either a cluster comprised of a large number of NBR1 is present, or a cluster is not present and only a much smaller number of NBR1 are present on the peroxisome surface. As a result of cluster formation, the total NBR1 on peroxisomes can greatly exceed the maximal (saturated) level of freely-diffusing NBR1. After only 1000 s, in Fig.~\ref{fig:figure1}(a), we observe more than an order of magnitude excess of NBR1 in the cluster (dotted red line) than in the highest surface concentration (dashed blue line).  NBR1 cluster formation provides a prompt all-or-none response that can be a mechanism for pexophagy selectivity. 

\subsubsection{Largest peroxisomes acquire NBR1 clusters first, and lose them last} 
The equations describing our quantitative model of NBR1 dynamics all depend on the peroxisome radius $R$ -- e.g. the rate at which NBR1 associates with the peroxisome surface; the critical concentration for cluster formation; and the change in cluster size in time. (See Eqns.~\ref{eq:jonnbr1}, \ref{eq:minimumc}, and \ref{eq:dynamical} below, respectively.) Because the NBR1 dynamics depend quantitatively on the peroxisome size, the behavior of NBR1 on a particular peroxisome, both freely-diffusing and cluster-bound, is affected by the radius of that particular peroxisome. They are also affected by the radii of other peroxisomes, since the NBR1 populations on all peroxisomes share the same cellular pool of NBR1.

In Fig.~\ref{fig:figure1}(b), there is a system of ten peroxisomes, with polydisperse radii between $R=0.05\mu$m and $R=0.50\mu$m. (Larger numbers of peroxisomes behave similarly, but are not as easily visualized. Each line in the figure corresponds to the number of NBR1 on a different individual peroxisome.) Initially, the freely-diffusing NBR1 on all ten peroxisomes increases approximately linearly, similar to the behavior of Fig.~\ref{fig:figure1}(a). Larger peroxisomes are able to recruit more NBR1, and so in Fig.~\ref{fig:figure1}(b) the largest peroxisome has the most NBR1, followed by the next largest peroxisome, and so on. Once the largest peroxisome reaches the critical concentration of freely-diffusing NBR1, soon after $t = 10$ s, it nucleates a cluster. This is followed by the next largest peroxisome nucleating a cluster, etc, until the five largest peroxisomes have a cluster, after which no further cluster nucleation occurs. The NBR1 pool not associated with peroxisomes is depleted as the nucleated NBR1 clusters grow and sequester NBR1; this reduction prevents smaller peroxisomes from reaching the critical NBR1 concentration and forming clusters. 

Fig.~\ref{fig:figure1}(b) shows that large peroxisomes form clusters, while small peroxisomes do not; and that clusters on larger peroxisomes grow earlier and shrink later than clusters on smaller peroxisomes. This implies that peroxisomes below a certain size will not have clusters. 

The clusters which have formed on the five largest peroxisomes then proceed to compete for material --- the smallest cluster, which is on the smallest peroxisome with a cluster, shrinks until it evaporates (bottom process in Fig.~\ref{fig:diagram}(b)), followed by the next smallest cluster, until there is only a single cluster remaining --- this is shown schematically in Fig.~\ref{fig:diagram}(c). This competition between clusters for NBR1, mediated by bulk diffusion between clusters, is known as Ostwald ripening \cite{yao93, voorhees85}. Only the larger drops retain clusters at late times \cite{brown15}.  

The initial nucleation of clusters is prompt and selects for larger peroxisomes. Furthermore, the initial nucleation of clusters is orders of magnitude faster than the slow resolution of clusters through Ostwald ripening dynamics. 

\subsubsection{Long lifetime for NBR1 clusters on even the smallest occupied peroxisomes} 
Fig.~\ref{fig:figure1}(c) examines how the smallest peroxisomal radius that has a cluster, $R_{clust}$, changes in time. We consider an exponentially distributed polydisperse distribution of peroxisome radii between $R=0.05$ $\mu$m and $R=0.50$ $\mu$m, using a system of 100 peroxisomes for good size resolution. We track the radius of the smallest peroxisome that has a cluster, $R_{clust}$, vs.\ time. Changes in $R_{clust}$ are due to the combined effect of nucleation of new clusters and evaporation of existing clusters.  $R_{clust}$ shows that large peroxisomes are selected for cluster formation and growth.  

At around $t=10$ s cluster nucleation begins with the largest peroxisomes, followed by nucleation on progressively smaller peroxisomes. Eventually cluster formation halts, even though not all peroxisomes are occupied by a cluster. As the larger peroxisomes formed clusters first, the end of cluster formation defines the radius of the smallest peroxisome occupied by a cluster at any time, $R^*$. Peroxisomes of radius $R^*$ and larger harbor a cluster at some time, while smaller peroxisomes are always unoccupied. Later, around $t=600$ s, clusters begin to evaporate from the smaller occupied peroxisomes --- and this ``Ostwald ripening'' causes $R_{clust}$ to slowly increase with time.

Each curve in Fig.~\ref{fig:figure1}(c) represents a different NBR1 concentration, as indicated in the legend. Higher NBR1 concentrations lead to earlier nucleation and evaporation, a lower $R_{clust}$ at all times, and a lower $R^*$.  In every case, we see that there is more than a decade in time where the smallest occupied peroxisome retains a cluster. During this period the size of that cluster is first growing and then shrinking.  Clusters on larger peroxisomes than $R^*$ survive even longer, as illustrated in Fig.~\ref{fig:figure1}(b).  

\subsection{Size selectivity and average peroxisome size}

Under our hypothesis that NBR1 clusters are required for pexophagy selectivity, the absence of NBR1 clusters on the smallest peroxisomes suggests some size-selectivity (determined by $R^*$). While our model does not directly address downstream pexophagy processes, if they are either fast enough to respond immediately to cluster formation (which progresses from the largest to the smallest peroxisomes) or slow enough to allow for evaporation of smaller NBR1 clusters (which progresses from the smallest to the largest peroxisomes) then size-selectivity could be even more substantial. 
 
We can explore some downstream effects of size-selectivity, using $R^*$. We explore three simplified cases: maximal, partial, or zero size selectivity. For maximum size selectivity all of the largest peroxisomes, defined as those with a radius greater than $R^\ast$, are degraded. Since larger peroxisomes are expected to harbor larger clusters (Fig.~\ref{fig:figure1}(b)), maximum size selectivity corresponds to the hypothesis that downstream processes first select peroxisomes with the largest NBR1 clusters.  For partial size selectivity, a fraction of all peroxisomes with radii greater than $R^*$ are degraded. This corresponds to the hypothesis that downstream processes randomly select peroxisomes with NBR1 clusters, but do not otherwise differentiate between them. For zero size selectivity, peroxisomes are selected randomly for degradation. This corresponds to the hypothesis that NBR1-clusters play no role in autophagy selectivity. 

A common measure of pexophagy activity is relative changes in the amount of catalase \cite{kim08, deosaran13, luiken92}, since catalase is an abundant peroxisomal marker. We will show that size-selectivity can significantly affect a volumetric measure such as total catalase fluorescence intensity, even when the number of degraded organelles remains unchanged.

\begin{figure}[tbp] 
	\centering
	\hspace{-1.0in}
	\begin{tabular}{cc}
		\hspace{-0.00in}\includegraphics[width=3.0in]{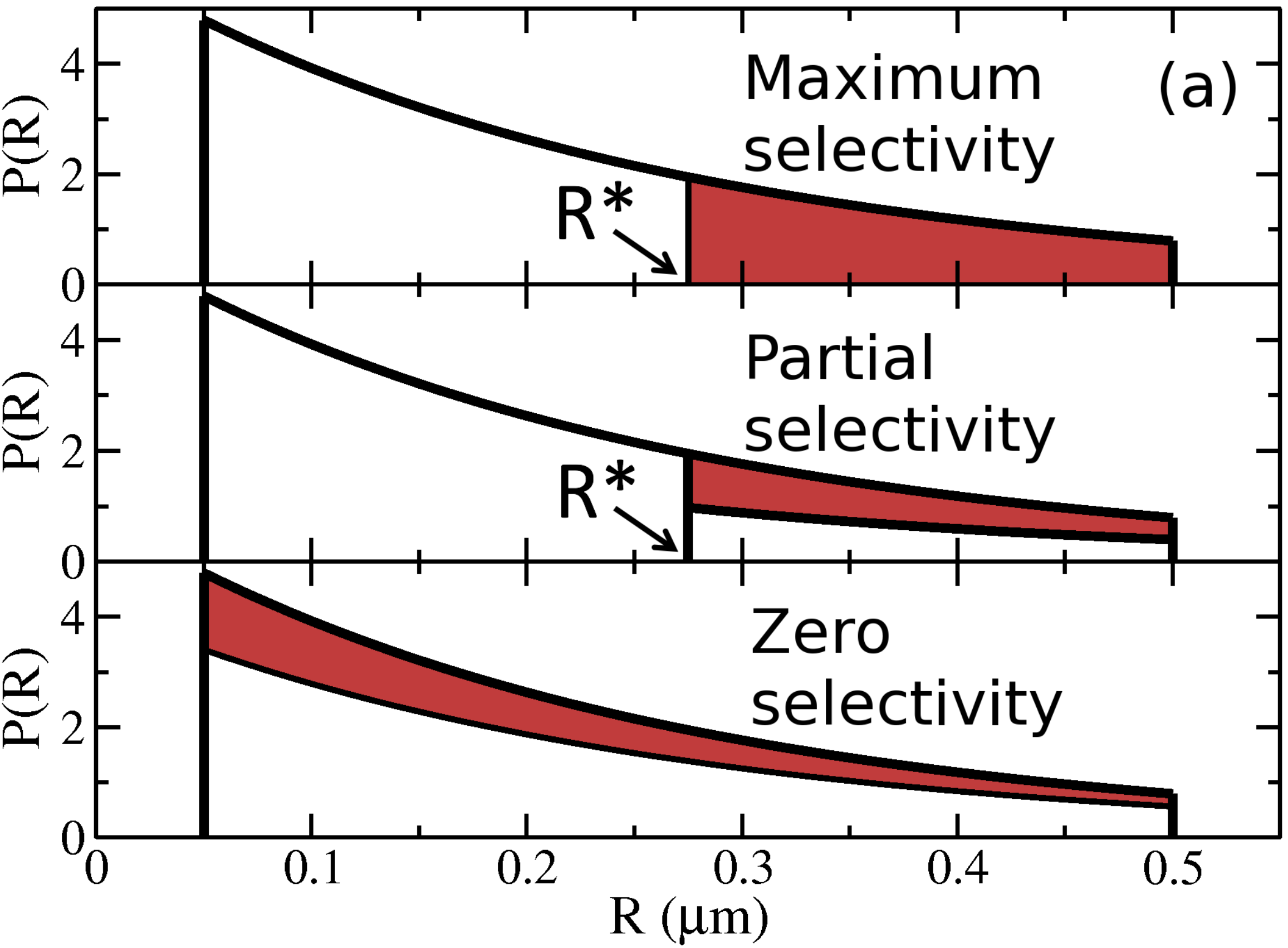} & 
		\hspace{-0.00in}\includegraphics[width=3.0in]{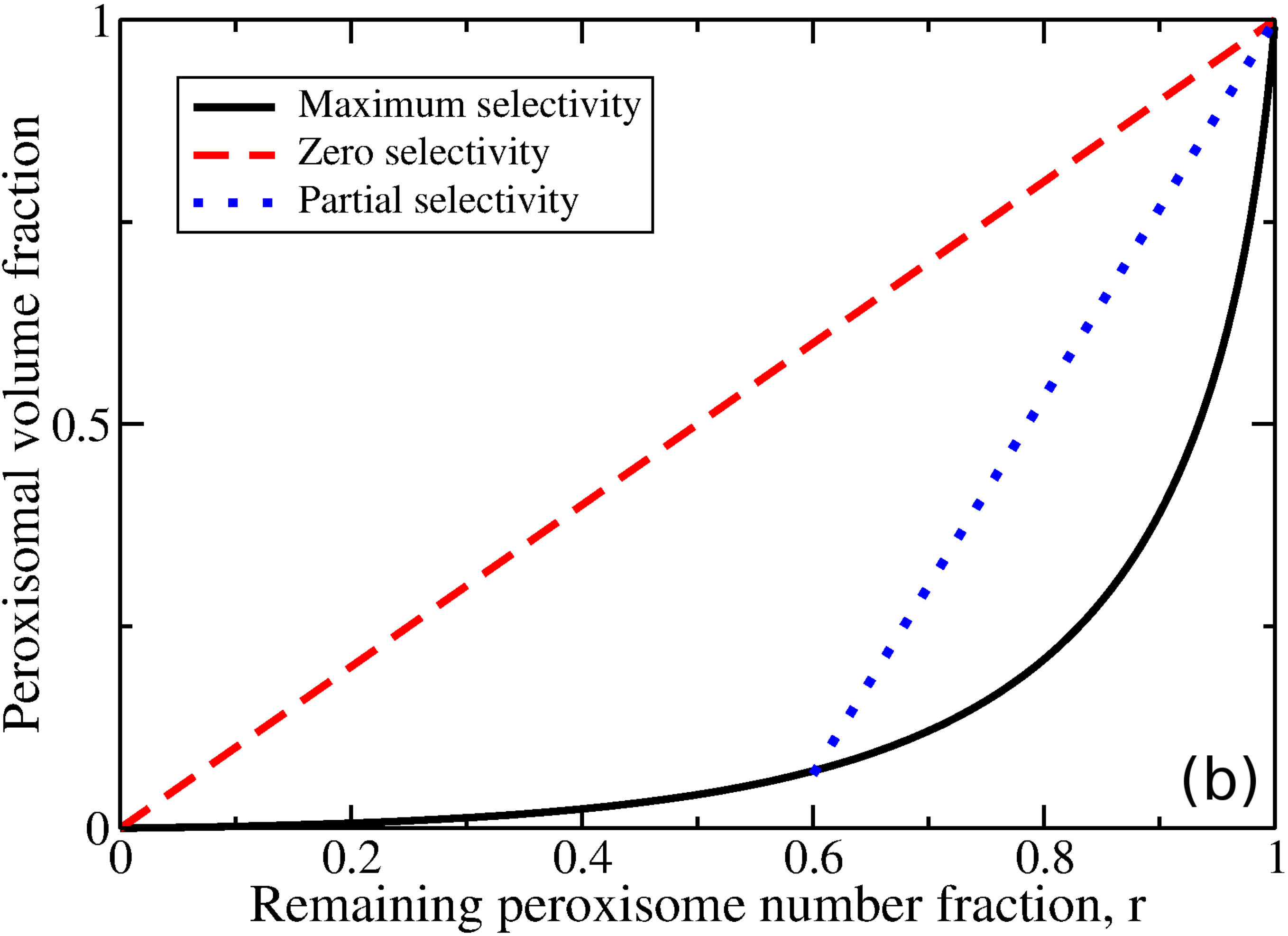}\\
	\end{tabular}
	\caption{ \label{fig:figure2} (a) Maximum size selectivity in pexophagy is achieved when all of the largest peroxisomes, of radius $R>R^*$, are degraded (shaded red area) from a distribution of peroxisome sizes (solid black line, illustrating an exponential distribution of abundance $P(R)$ vs radius $R$). Partial size selectivity is achieved when a fraction $p$ of the larger peroxisomes, with $R>R^\ast$, are degraded. Zero size selectivity is achieved when peroxisomes are randomly chosen for degradation, independent of size. (b) The remaining peroxisomal volume fraction vs.\ the  remaining number fraction $r$, after various amounts of zero selectivity degradation (dashed diagonal red line) or maximal selectivity degradation (solid black line, given by Eq.~\ref{eq:vmax}). Partial selectivity is equivalent to maximum selectivity a fraction $p$ of the time, and so lies on a line interpolating between maximum selectivity with the same $R^\ast$ and no autophagy (at $r=1$). One such interpolating line (dotted blue) is illustrated.}
\end{figure}

\subsubsection{Size selectivity significantly affects catalase abundance} 
Maximum, partial, and zero size selectivity are illustrated in Fig.~\ref{fig:figure2}(a) for exponentially distributed peroxisomal radii between $R_{min}=0.05$ $\mu$m and $R_{max}=0.5$ $\mu$m. The red shaded region corresponds to a fixed fraction of degraded peroxisomes under the three scenarios of maximal, partial, or zero selectivity (as indicated). For maximum size selectivity, all peroxisomes with radii greater than $R^\ast$ are degraded. For partial size selectivity, only peroxisomes with radii greater than $R^*$ are degraded, however only a fraction $p$ of the peroxisomes in this size range are randomly degraded. For zero size selectivity, peroxisomes are selected for degradation randomly, so that peroxisomes of different size are selected in proportion to their abundance. Maximum and zero selectivity correspond to limiting cases of partial selectivity, with $p=1$ or $R^*=R_{min}$, respectively. 

We consider a scenario where a certain fraction of the total number of peroxisomes are degraded, with a fraction $r$ remaining. For the same number of peroxisomes removed, maximum, partial, and zero size selectivity will each result in a different fraction of the total peroxisomal volume being removed upon degradation. We calculate the volume fraction remaining as peroxisomes are degraded (shown in Computational Methods), and the results of these calculations are shown in Fig.~\ref{fig:figure2}(b). For zero size selectivity, the remaining peroxisomal volume is strictly proportional to the remaining peroxisomal number (dashed red line). For maximum selectivity, the remaining volume fraction decreases sharply as the number fraction is decreased (solid black line). This is  because only the largest peroxisomes are targeted when size-selectivity is maximal. Partial selectivity (dotted blue line) interpolates between maximum selectivity and no autophagy (at $r=1$). Only one interpolating line is shown for illustrative purposes. 

We see that the reduction in peroxisome volume, and the corresponding reduction in catalase intensity, is a combination of the number of peroxisomes removed and the size selection of those peroxisomes. Many more peroxisomes need to be removed with zero selectivity to match the same volume reduction at maximum selectivity.  Since a change of catalase intensity is a proxy for a change in peroxisomal volume, our results indicate that catalase intensity can be affected by either changing the number of peroxisomes degraded {\em or} by changing the size-selectivity. 

\subsection{p62 model description}

Experimental p62 inhibition with siRNA causes an increase in the catalase signal \cite{deosaran13}. This relative increase of catalase has been previously interpreted as a decrease in the pexophagy rate due to p62 inhibition \cite{deosaran13}, but any change in size-selectivity could confound this interpretation. To explore this possibility, we now extend our model to consider how p62 could affect size selectivity of peroxisomes. The results of our p62 model development are presented here without mathematical details (see details below in Computational Methods).  Our model for p62 dynamics has two aspects: p62 association with and recruitment to peroxisomes, and p62 inhibition of NBR1 clusters.

Similar to NBR1, p62 can in principle bind to ubiquitin associated with peroxisome membranes using its UBA domain \cite{long08, raasi05}. However (details in Computational Methods), the dissociation constant is too large to expect significant p62 binding to ubiquitin. Nevertheless, the p62 PB1 domain has a strong affinity to other PB1 domains \cite{wilson03}, and so p62 can bind to membrane-associated NBR1 through its PB1 domain. Thereafter, associated p62 can form filaments through PB1-PB1 interactions \cite{lamark03,bienz14}.

p62 recruitment to NBR1 on the peroxisome surface is modeled as diffusion-limited arrival to and association with NBR1 on the peroxisome membrane, or to p62 already associated with NBR1. These are both quantitatively described by a standard equation for diffusion-limited association to absorbing targets on a sphere (Eqn.~\ref{eq:jonp62} below).

Polymer physics leads us to hypothesize that polymeric chains of p62 associated with NBR1 could reduce NBR1 self-association through steric repulsion. Such a repulsion arises as the entropic contribution to the free-energy of the polymers is decreased when brought close together \cite{hristova94}. Such steric repulsion of membrane associated proteins can prevent growth of protein clusters \cite{sieber07}, can lead to cluster segregation \cite{sens04}, and can even inhibit phase separation of associated lipids \cite{scheve13}. To explore the consequences of a strong steric repulsion between p62-associated NBR1, our model does not allow NBR1 that is associated with p62 to participate in cluster formation or growth.

\subsection{p62 Results}

\begin{figure}[tbp] 
	\centering
	\hspace{-1.00in}
	\begin{tabular}{cc}    
		 \hspace{-0.00in}\includegraphics[width=3.0in]{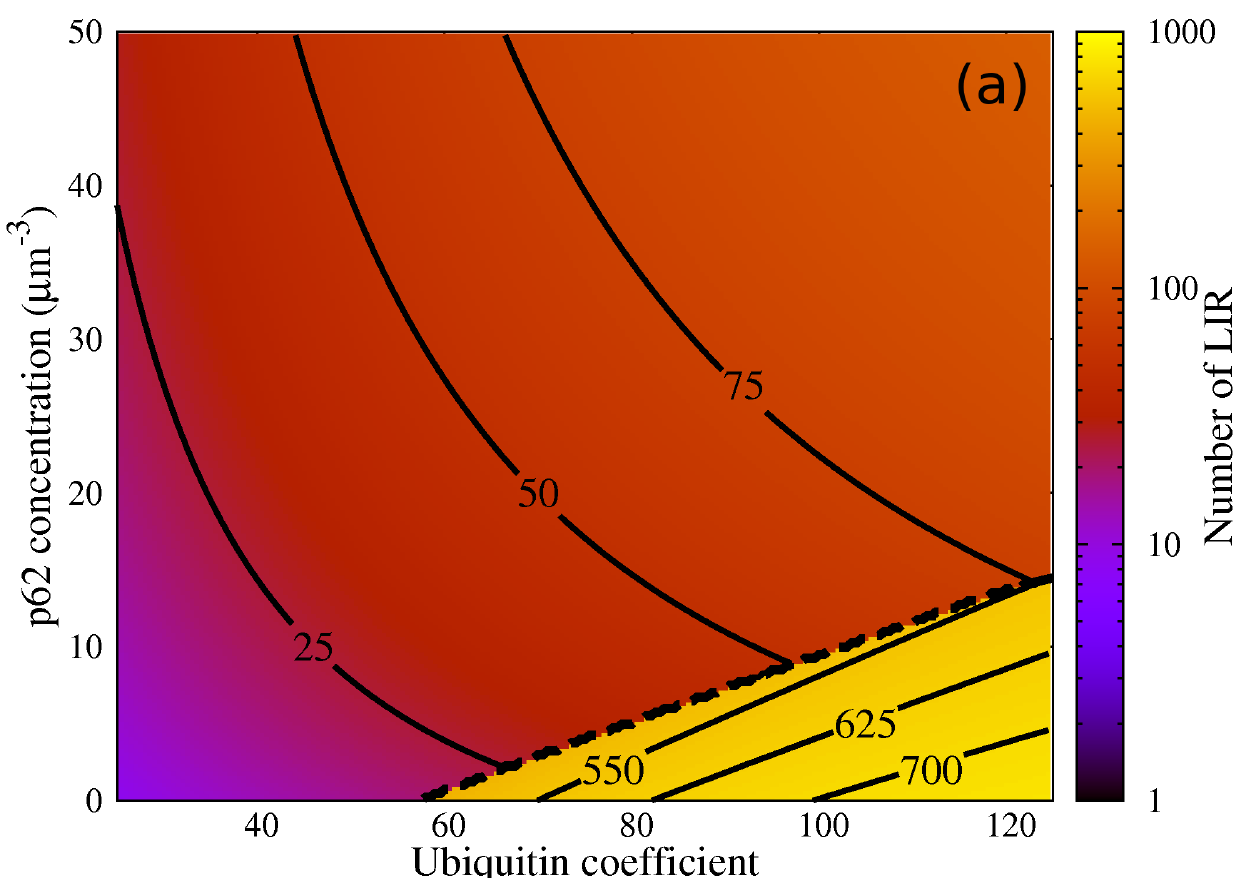} &   
		\hspace{-0.00in}\includegraphics[width=3.0in]{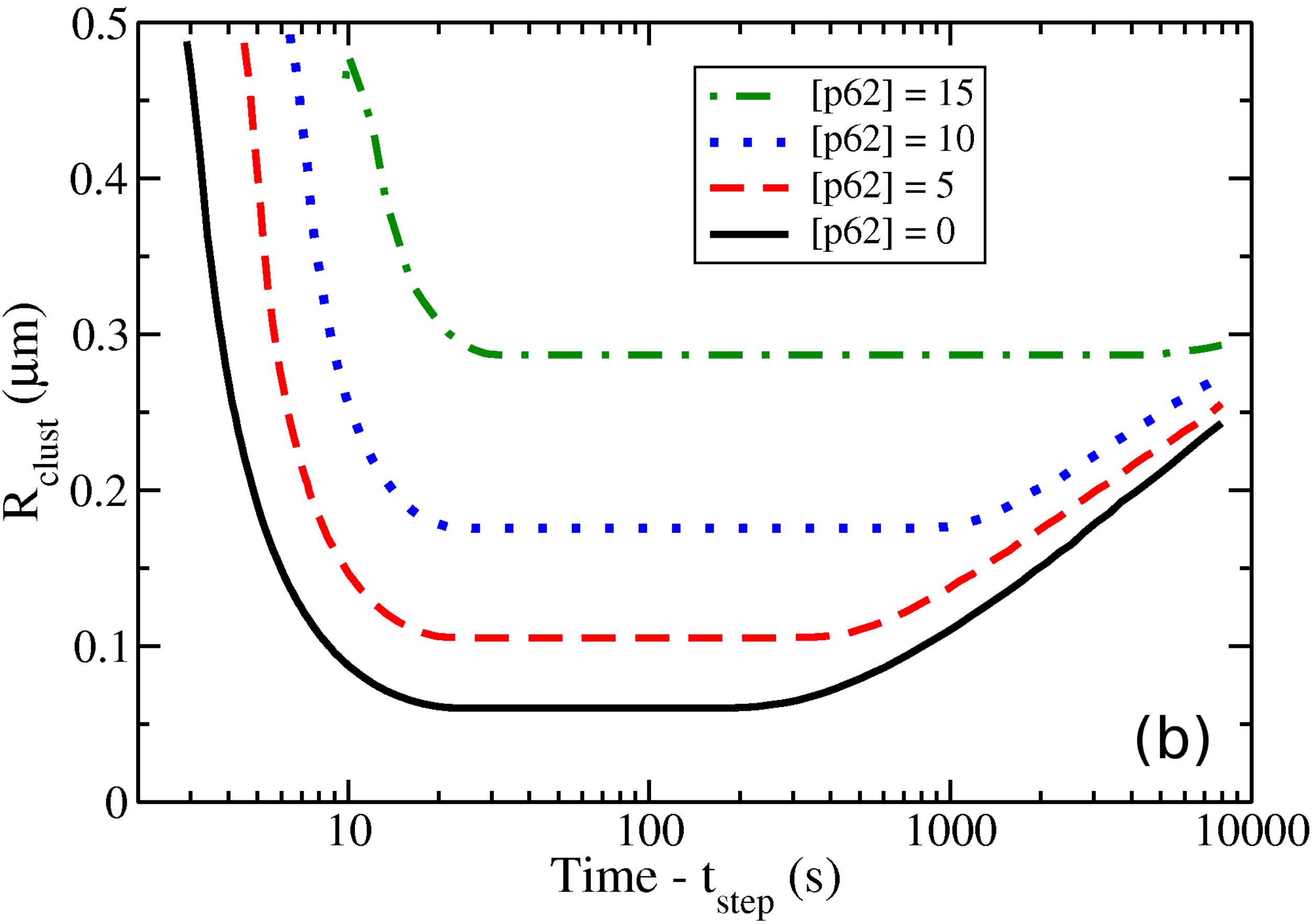} \\
		\hspace{-0.00in}\includegraphics[width=3.0in]{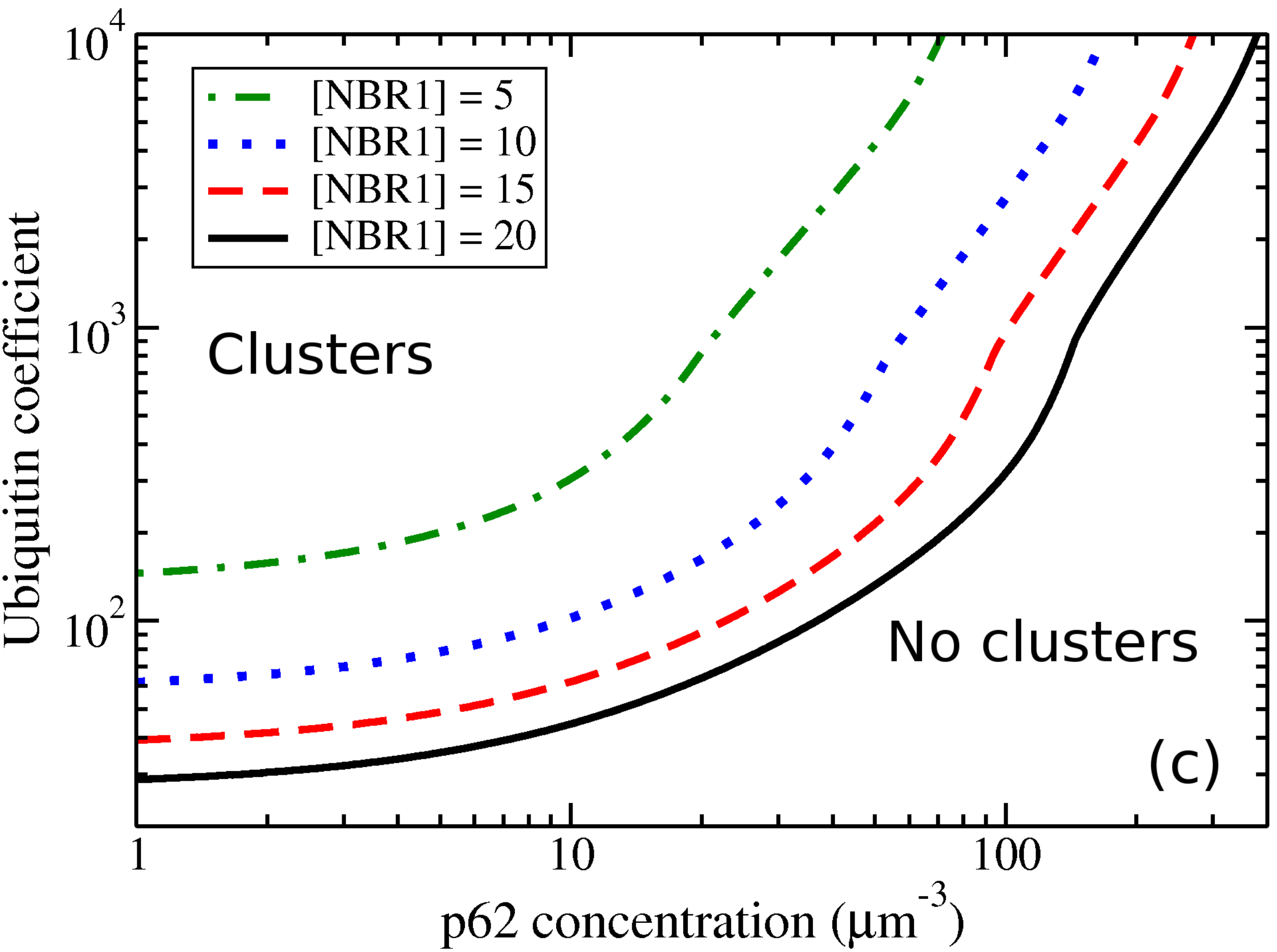} &   
		 \hspace{-0.00in}\includegraphics[width=3.0in]{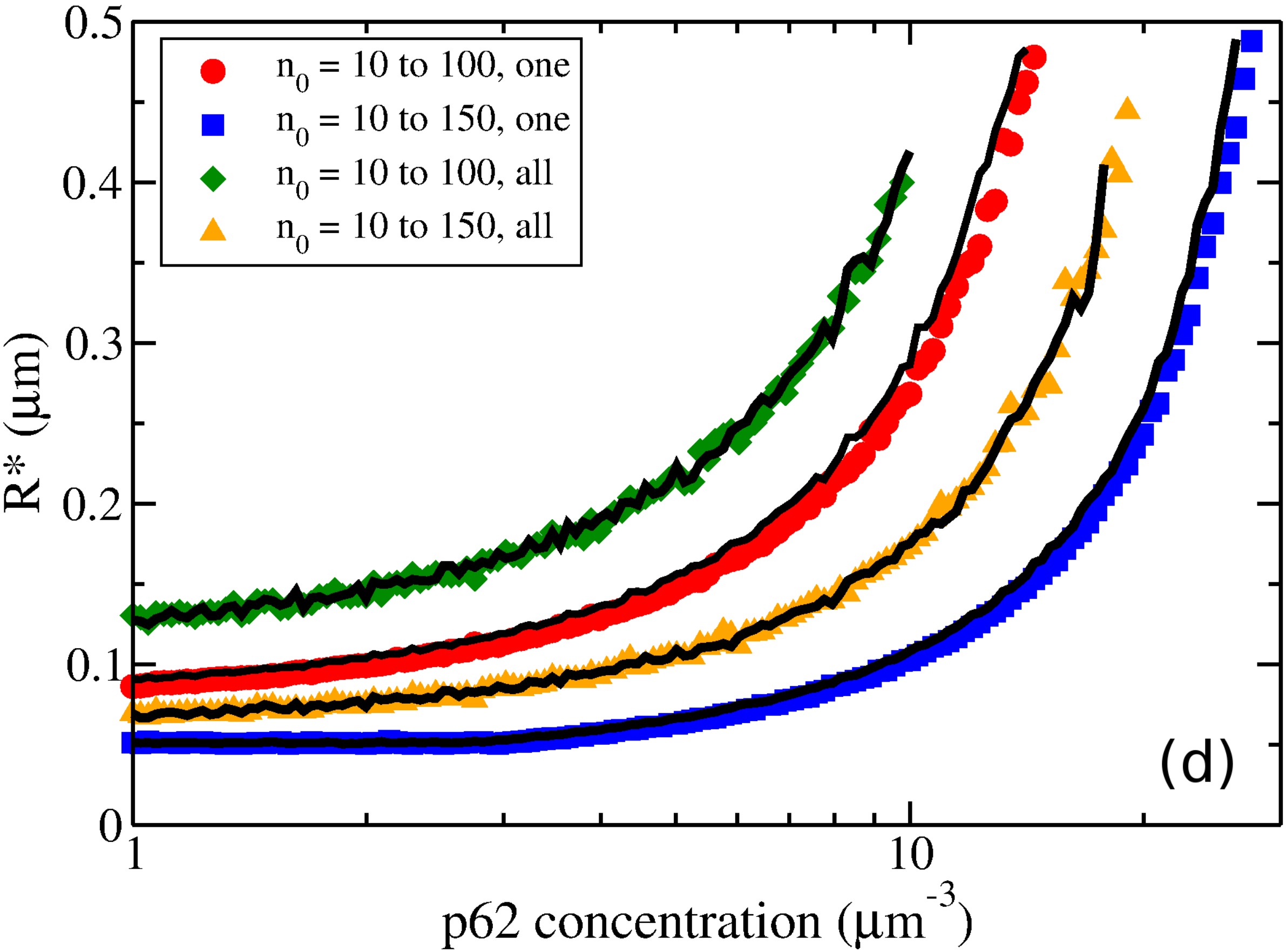}  
	\end{tabular}
	\caption{\label{fig:figure3}  (a) The steady-state number of LIR domains on a single peroxisome after equilibration, given by the sum of the number of NBR1 and p62, as cellular [p62] (in $\mu$m$^{-3}$) and number of ubiquitin are varied.  $R=0.25$ $\mu$m, [NBR1] = 10 $\mu$m$^{-3}$, and $V = 100$ $\mu$m$^3$.  NBR1 clusters are present below the dashed line. Increasing ubiquitin increases the LIR count, and so does increasing [p62] when no clusters are present.  However, increasing [p62] suppresses cluster formation and decreases the LIR count within the clustering regime. 
	(b) The radius of the smallest peroxisome with a cluster, $R_{clust}$, vs.\ time, following induction of cluster formation after increase of the ubiquitin coefficient at $t=t_{step}$. [p62] is varied as indicated, and we otherwise use the same conditions as in Fig.~\ref{fig:figure1}(c).  Initially the largest peroxisomes form clusters, then for a decade of time no more clusters are formed, followed by cluster evaporation from the smallest occupied peroxisomes. Increasing [p62] leads to more selective cluster formation, with only the larger peroxisomes occupied by clusters.  
		(c)  Lines indicate the boundary between regimes with at least one NBR1 cluster on some peroxisome (above) and those with no clustering on any peroxisome (below), as both cellular [p62] and ubiquitin are varied as indicated. Different lines correspond to cellular [NBR1] as indicated by the legend. Other conditions are the same as in Fig.~\ref{fig:figure1}(c). For higher ubiquitin, lower [p62], or higher [NBR1] we observe clusters. Even for higher [p62], sufficiently high ubiquitin will still lead to NBR1 clusters -- though note the logarithmic scale.  
		(d) After the ubiquitin coefficient $n_0$ is suddenly raised (as indicated by the legend), the data shows the radius of the smallest peroxisome with a cluster at any time, $R^\ast$, vs.\ cellular p62 concentration. The disuse and damage scenarios correspond to ubiquitin being increased on all peroxisomes (``all", diamonds and triangles) or increased only on a single peroxisomes (``one", circles and squares), respectively. The damage scenario leads to significantly smaller $R^\ast$, as does a larger final ubiquitin coefficient. While the data points represent an initial ubiquitin coefficient of $n_0=10$,  the black lines represent initial ubiquitin coefficient of $n_0=25$, and their agreement indicates that the initial ubiquitin coefficient is not significant. The number of ubiquitin on a peroxisome is $N_{Ub}=n_0(R/R_0)^2$, with $R_0=0.25$ $\mu$m.}
\end{figure}

In our quantitative model a p62 filament on a membrane-associated NBR1  prevents that NBR1 from participating in cluster formation or growth -- which should reduce NBR1 numbers. Despite this cluster-inhibition mechanism, NBR1 and p62 both contain a LIR domain, which interacts with the machinery for the formation of autophagosomes \cite{lin13, birgisdottir13}. Given these opposing effects, how does the number of membrane-associated LIR domains change as the p62 concentration is varied? How does greater or lesser amounts of surface ubiquitination affect the function of p62? 

\subsubsection{p62 inhibits NBR1 clusters} 
For an individual peroxisome, we show in Fig.~\ref{fig:figure3}(a) how the total steady-state number of LIR on a peroxisome changes as both the global p62 concentration and the peroxisomal number of ubiquitin are varied. For larger p62 concentration (the purple/red region) NBR1 clusters are suppressed, dramatically reducing LIR content.  Within this non-clustering regime the LIR content slowly increases with increasing p62. For smaller p62 concentration (the yellow region) we see a striking increase in the LIR count due to formation of an NBR1 cluster. The clustering regime has approximately 10$\times$ the number of LIR domains as the non-clustering regime.  Within the clustering regime, increasing p62 {\em decreases} the LIR count by decreasing the cluster size. 

\subsubsection{p62 inhibition of NBR1 clusters enhances size selectivity}  
In Fig.~\ref{fig:figure1}(c) we showed how there is a time-dependent threshold peroxisome size $R_{clust}$ - below this size the peroxisomes do not have a cluster, and above this size peroxisomes harbour a cluster. Here we investigate how p62 can change this threshold peroxisome size.

In Fig.~\ref{fig:figure3}(b) we show the radius of the smallest peroxisome with an NBR1 cluster, $R_{clust}$, vs.\ elapsed time after the ubiquitin level is simultaneously increased on all peroxisomes. (We use ubiquitin increase to initiate autophagy here since it recruits both NBR1 and p62.) We use the same radius distribution as Fig.~\ref{fig:figure1}(c). The ubiquitin increase induces NBR1 uptake to peroxisome surfaces.  We consider various p62 concentrations, as indicated by the legend. Increasing the p62 concentration causes an increase in the size of the smallest peroxisome with a cluster, and limits NBR1 clusters to larger peroxisomes. If only peroxisomes with clusters are degraded by autophagy, p62 increases the threshold size for peroxisome degradation, and thus enhances size selectivity.

\subsubsection{Sufficient ubiquitin allows NBR1 clusters to form on larger peroxisomes despite p62} 
In Fig.~\ref{fig:figure3}(c) we show that sufficiently high p62 concentrations (or low ubiquitin coverage) can completely inhibit clusters on all peroxisomes. Each curve indicates the minimum ubiquitin coefficient that leads to NBR1 cluster formation on any peroxisome, under the same radius distribution as Fig.~\ref{fig:figure1}(c).  Below the line, no NBR1 clusters form, while above the line there is at least one NBR1 cluster. The different lines, as indicated by the legend, correspond to different cellular NBR1 concentrations. We see that increasing NBR1 leads to more peroxisomes with clusters. 

While sufficiently high ubiquitination allows NBR1 cluster formation at any given p62 concentration, the clusters still only form on larger peroxisomes. This is shown in Fig.~\ref{fig:figure3}(d), where the radius of the smallest peroxisome with an NBR1 cluster, $R_{clust}$, vs.\ [p62] is shown. With more p62, the radius of the smallest peroxisome occupied by a cluster increases. 

\subsubsection{Single organelle damage has weaker size selectivity than multi-organelle disuse} 
Fig.~\ref{fig:figure3}(b) shows how increasing the p62 concentration restricts NBR1 clusters to larger peroxisomes. For each curve in Fig.~\ref{fig:figure3}(b), the radius of the smallest peroxisome with a cluster reaches a minimum, $R^*$. This is the radius of the smallest peroxisome that ever has a cluster.

In Fig.~\ref{fig:figure3}(d) we see that $R^\ast$ increases as the p62 concentration increases. We consider two values of the initial ubiquitin coefficient ($n_0=10$ with colored points as indicated by the legend and $n_0=25$ with corresponding thin black lines) and two values of the final ubiquitin coefficient ($n_0=100$ with circles and diamonds, or $150$ with squares and triangles, as indicated in the legend). We suddenly change the ubiquitin number on peroxisomes from the low initial  value, which does not support clustering, to a high final value that does support clustering. The agreement between points and black lines indicates that the {\em initial} ubiquitin coefficient does not significantly affect the clustering. However, the final ubiquitin coefficient does affect clustering.

When we compare the minimum peroxisome size with a cluster, $R^*$, between a ``disuse'' scenario (shown schematically in Fig.~\ref{fig:diagram}(d)), where {\em all} peroxisomes increase their ubiquitin (green diamonds and orange triangles in Fig.~\ref{fig:figure3}(d)), and a ``damage'' scenario (Fig.~\ref{fig:diagram}(e)), where only {\em one} peroxisome increases its ubiquitin (red circles and blue squares in Fig.~\ref{fig:figure3}(d)) we see that for damage significantly smaller peroxisomes will form NBR1 clusters. Hence the damage scenario decreases size selectivity, allowing more damaged peroxisomes to be selected with moderate amounts of ubiquitin. 

\section{Discussion}

\subsection{Summary} 
Autophagy selectivity amounts to an all-or-none response, where each substrate is either selected or not selected.  Selectivity is thought to be mediated by autophagy receptor proteins.   Some autophagy receptor proteins can be observed to form microdomains on substrates \cite{rogov14, cemma11, wild11, mostowy11}. Our primary hypothesis is that the presence or absence of receptor clusters on individual organelles provides a necessary all-or-none response for autophagy selectivity, for at least some types of organelles. 

There can be hundreds of peroxisomes in a single mammalian cell, and pexophagy can result from either damage of individual peroxisomes or from more generic deproliferation  \cite{ezaki11}. The signaling protein ubiquitin and the autophagy receptor proteins NBR1 and p62 participate in pexophagy \cite{kim08, deosaran13, brown14}. NBR1 is essential for pexophagy; also essential are the specific domains of NBR1 that govern NBR1 association with itself, with plasma membranes, with ubiquitin, and with autophagosomes. We explore our primary hypothesis by quantitatively modeling the dynamics of NBR1 cluster formation on individual peroxisomes. 

With sufficient cellular NBR1 and peroxisomal ubiquitin, our model leads to the formation and growth of NBR1 clusters. Clusters are found to form rapidly on the largest peroxisomes, and subsequently on smaller ones. After cluster nucleation, competitive Ostwald ripening progressively removes initial clusters --- starting from the smaller peroxisomes. The smallest peroxisomes (below a critical radius $R^\ast$) never form clusters. Thus we find that NBR1 clustering is strongly size-selective --- NBR1 clusters are found on large peroxisomes and not on small peroxisomes. Nevertheless, higher NBR1 concentrations or peroxisomal ubiquitin levels can reduce size selection by reducing $R^\ast$.

While p62 is not essential for pexophagy, experiments show that lower p62 levels increase the cellular abundance of peroxisomal catalase -- indicating a significant role in pexophagy \cite{deosaran13}. We echo our primary hypothesis with a secondary hypothesis that p62 influences pexophagy by affecting NBR1 cluster formation. Specifically, we hypothesize that p62 association with NBR1 and subsequent p62 filament formation will sterically hinder and {\em inhibit} NBR1-NBR1 association.  By modelling this hypothesis, we find that p62 can significantly affect size-selectivity and that this alone may be sufficient to explain the catalase response of p62 inhibition.

\subsection{NBR1 clusters and size-selectivity} 
Deosaran \emph{et al} \cite{deosaran13} introduced the idea that a `critical mass' of autophagy receptor proteins is necessary to target peroxisomes to autophagosomes. Supporting this idea, when NBR1 exceeds a critical concentration on the peroxisomal surface nucleation of an NBR1 cluster is followed by a sudden and localized further increase in LIR numbers, and provides an all-or-none signal on individual peroxisomes. 

We found that cluster formation favours larger peroxisomes because of a lower surface concentration threshold for cluster formation. This is seen in Figs.~\ref{fig:figure1}(b) and \ref{fig:figure1}(c), with the first cluster forming on the largest peroxisome and then progressing towards smaller peroxisomes. Subsequent cluster evaporation begins on the smallest peroxisome with a cluster (peroxisomal radius $R^\ast$) and progresses towards larger peroxisomes \cite{brown15}. NBR1 clustering provides a consistent signal for autophagy to target and degrade large peroxisomes. This size-selectivity of cluster formation and evaporation, in combination with our hypothesis that NBR1 clusters are an essential part of pexophagy selectivity, is consistent with early reports that larger peroxisomes are preferentially degraded by pexophagy \cite{veenhuis78} and that the required proteins for pexophagy depend on peroxisome size \cite{nazarko09}. 

Cluster formation occurs very quickly, with the number of clusters peaking after an increase in the ubiquitin level in $\lesssim30$ s in Fig.~\ref{fig:figure3}(b). Following cluster formation, evaporation of clusters is slower, beginning after $10^2$-$10^4$ s in Fig.~\ref{fig:figure3}(b). This evaporation is nevertheless faster than reported for pre-existing clusters \cite{brown15} because some clusters are nucleated close to the threshold for evaporation. These cluster formation and evaporation timescales suggest that the selection of peroxisomes by NBR1 cluster formation could occur well within the timescale of mammalian or yeast pexophagy, which take days \cite{moody76, iwata06} or hours \cite{platta07, veenhuis83}, respectively.

While we hypothesize that NBR1 receptor clusters are necessary for autophagy selectivity, they may not be sufficient. Our scenario of partial selectivity,  illustrated in Fig.~\ref{fig:figure2}(a), reflects this possibility.

While NBR1 clusters on peroxisomes and their preference for larger peroxisomes appear to be viable selectivity mechanisms according to our quantitative modeling, they remain hypotheses. We do note that organelle-size dependent activity arising from cluster size modifications has been previously proposed in the context of membrane recycling \cite{vagne15}. Moreover, our proposed mechanism may apply for mitochondria in a cellular anti-viral response mediated by MAVS (also known as CARD or IPS) proteins \cite{odendall14}. Before viral infection, MAVS are not activated, and are spread out on the mitochondrial membrane \cite{hou11} and are evenly distributed between mitochondria \cite{onoguchi10}. Upon viral infection MAVS are activated, and subsequently cluster on the mitochondrial membrane \cite{hou11, xu14}. Significantly, some mitochondria have significant MAVS while other mitochondria have little to no MAVS \cite{onoguchi10}. Limited data suggests that large mitochondria retain MAVS while small mitochondria do not \cite{onoguchi10}.  The MAVS anti-viral phenomenology appears qualitatively similar to our NBR1 model, and so provides some support for our model behavior.  

\subsection{Pexophagy selectivity: disuse vs damage} 
In our model an increase in ubiquitin number acts as an initial signal that can lead to additional NBR1 and p62 accumulation, possibly followed by autophagic degradation. Ubiquitination of peroxisomes has been shown sufficient to induce pexophagy \cite{kim08}; ubiquitin is thought to recruit NBR1, the primary autophagy receptor protein for peroxisomes, to the peroxisomes membrane \cite{deosaran13}; and earlier modeling suggests that an increase in ubiquitin could be a natural and self-correcting response of the cell to requiring fewer peroxisomes \cite{brown14}. Ubiquitin is known to play a role in the routine import of peroxisome matrix proteins \cite{brown14}, and is part of the quality control system for damaged proteins on peroxisomes \cite{platta04, kiel05, erdmann05} that is distinct from the well known role of ubiquitin as a signal for the ubiquitin-proteasome system \cite{kraft10}.

In Fig.~\ref{fig:figure3}(d) we explored two extreme cases of increases in the ubiquitin level: a global increase, where ubiquitin levels on all peroxisomes increase, and an increase of the ubiquitin level on a single peroxisome. A global increase (shown schematically in Fig.~\ref{fig:diagram}(d)) could be due to the removal of peroxisome proliferators \cite{hess65} or a change in growth medium \cite{platta07}, both of which can result in a decrease in peroxisome numbers as superfluous peroxisomes are degraded. An increase on a single peroxisome (shown schematically in Fig.~\ref{fig:diagram}(e)) could be due to damage \cite{shibata13, zutphen11}.

For a disuse scenario, size-selectivity would lead to the expectation that larger peroxisomes would be preferentially degraded until the decrease in peroxisome numbers had been achieved and ubiquitin levels reduced \cite{brown14} so that NBR1 clusters are no longer formed. This is consistent with observations that larger peroxisomes are preferentially degraded when reducing peroxisome numbers \cite{veenhuis78}.  

For the damage scenario, there is little competition for the NBR1 needed to form clusters since only a few peroxisomes have an elevated ubiquitin level. This results in less size-selection (lower $R^\ast$) than in the disuse scenario, as seen in Fig.~\ref{fig:figure3}(d). Nevertheless, even in the damage scenario  larger peroxisomes more easily form NBR1 clusters, suggesting that larger peroxisomes may need less damage to be selected for autophagy. 

\subsection{Selectivity with p62} 
The autophagy receptor protein p62 is important for pexophagy \cite{kim08}. Inhibition of p62 with siRNA increases total peroxisomal catalase   \cite{deosaran13}.  We have demonstrated that when p62 inhibits NBR1 clustering in our model, increased p62 levels cause greater size-selectivity: the size threshold above which peroxisomes have NBR1 clusters is pushed to larger peroxisomes by increased p62 concentrations. This in itself would be sufficient to explain the catalase increase following p62 inhibition. Volumetric measures of autophagy such as catalase will report the combination of number and volume. To assess the number of organelles targeted by autophagy, the number should be directly assessed.

\subsection{Experimental signatures of NBR1 clusters} 
(a) The most striking result of our hypothesis, that NBR1 clusters are necessary for downstream degradation by the autophagy system, is significant size-selectivity. Large peroxisomes will be preferentially degraded over small peroxisomes. One way of measuring this effect would be to measure the fluorescence intensity of a tagged peroxisomal protein, such as catalase, together with the degree of colocalization with a protein associated with autophagosomes, such as LC3 \cite{hansen11}. Our model results indicate that peroxisomes with significant colocalization would have a larger average catalase intensity compared to peroxisomes with little or no colocalization. 

(b) Our model also indicates that formation of NBR1 clusters will significantly affect the number of NBR1 associated with peroxisomes of similar size. For the parameters of Fig.~\ref{fig:figure3}(a), the differences are approximately ten-fold. While surface ubiquitin concentrations may differ between peroxisomes of similar sizes, and so could determine which peroxisomes have clusters, ubiquitin alone appears unlikely to be able to directly affect non-cluster NBR1 to the same extent \cite{brown14}.

With NBR1 clustering, we have shown that peroxisomes will either have a cluster and have a large amount of NBR1, or not have a cluster and have a small amount of NBR1. After pexophagy is induced, e.g. by the removal of peroxisome proliferators, the fluorescence of NBR1 colocalizing with catalase should therefore have a bimodal distribution. Qualitatively, the all-or-none colocalization of NBR1 with peroxisomes suggested by, e.g., Fig.~5 of \cite{deosaran13} is consistent with our model.

With time-resolved imaging, NBR1 localization to individual peroxisomes is expected to be similar to Fig.~\ref{fig:figure1}(b). The signature of cluster formation would be a sudden increase of NBR1 number on peroxisomes with clusters at the same time as NBR1 numbers gradually decrease on peroxisomes without clusters.

(c) Our secondary hypothesis is that p62 inhibits NBR1 cluster formation.  If true, we would not expect p62 to have a bimodal distribution since p62 would decorate the freely-diffusing NBR1 (which does not change with cluster formation), but not the NBR1 in clusters (which does).  We caution that this lack of bimodality may only apply at the early stages of substrate selection, due to the many cellular roles of p62.

If p62 inhibits NBR1 clustering, then when p62 expression is knocked down by siRNA, there should then be an {\em increase} in the number of NBR1 clusters on peroxisomes.

\subsection{Important challenges of selective autophagy} 
We have argued that  selective autophagy requires an `all or none' signal to decide whether or not to degrade a substrate. We have proposed a receptor clustering mechanism for this all or none signal for peroxisomes, which leads to a prediction that large substrates are more likely to be selected for degradation. This is consistent with some reports of a preference for degradation of large peroxisomes. However, larger peroxisomes might simply be older and/or more damaged. Distinguishing damage-induced selective autophagy from, e.g., disuse-induced selective autophagy is an interesting challenge. Our suggestion that damage-induced selective autophagy should be {\em less} size-selective may help in this regard.  

Although our cluster selectivity hypothesis does not uniquely explain size-selectivity, our model does present a working hypothesis for the basic mechanism for substrate selection in selective autophagy. Furthermore, the long lifetime of receptor clusters (see Figs. 1(c) or 3(b)) allows ample time for downstream processes to positively recognize the all or none signal provided by the receptor clusters, and to avoid `false-positive' triggers on e.g. stochastic fluctuations. Such a stable all-or-none signal is a challenge both for time-resolved microscopy studies of receptor dynamics, but also a challenge for any competing models of the selectivity mechanism of selective autophagy that arise in the future. 

\section{Computational methods}

\subsection{NBR1 and p62 structure}
Each NBR1 molecule contains several regions that are essential for pexophagy. The LC3-interacting region (LIR) interacts with the proteins of the autophagy system \cite{lin13,birgisdottir13}. The UBA region can bind to ubiquitin \cite{lin13,vadlamudi96}, and allows attachment to ubiquitin-tagged substrates \cite{kraft10}. The Phox and Bem1p (PB1) region can bind PB1 regions on other proteins \cite{lamark03}, and coiled-coil regions promote self-interaction \cite{deosaran13}. The distinctive `J' region allows NBR1 to anchor to membranes \cite{deosaran13}.

Similarly, p62 also has LIR, UBA, and PB1 regions \cite{lin13,birgisdottir13,kraft10,lamark03}. Distinctively, the PB1 region of p62 can bind two other PB1 regions, forming chains of p62 \cite{ciuffa15}, unlike NBR1 which can only bind one other PB1 \cite{lamark03}. Unlike NBR1, p62 has no J region.

\subsection{NBR1 and p62 association with peroxisomes and each other}
NBR1 and p62 both contain ubiquitin-interacting UBA regions \cite{vadlamudi96,walinda14}. However, these UBA regions have relatively weak affinities compared to expected cellular abundances.  For NBR1, $K_{d,UBA} =3- 4 \mu$M \cite{walinda14} while typical abundances in human cell lines are no more than $\approx 125$ ppm \cite{geiger12}, or a concentration [NBR1] $\lesssim$ 0.6 $\mu$M since $1$ ppm corresponds to approximately $5$ nM \cite{milo13}. For p62, $K_{d,UBA} = 540-750$ $\mu$M \cite{long08, raasi05} with typical abundances in human cell lines no more than $\approx 300$ ppm, or  [p62] $\lesssim 1.5$ $\mu$M.  While phosphorylation of p62 significantly increases polyubiquitin association, the enhancement appears to be no more than three-fold \cite{matsumoto11}. Phosphorylation decreases the association of NBR1 with ubiquitin \cite{nicot14}. The relatively weak affinities of NBR1 and p62 UBA regions with ubiquitin implies that there should only be a small  fraction of these receptors on surface-displayed ubiquitin.  

How can NBR1 significantly associate with peroxisomes, if not by association with ubiquitin? The essential J region of NBR1  mediates membrane association even without ubiquitin \cite{mardakheh10}.  $K_d$ of typical amphipathic helices can be as low as $20$ nM \cite{stahelin03}. Coincidence of ubiquitin and membrane association \cite{carlton05, deosaran13} could further decrease the effective $K_d$ of membrane association through an enhanced on rate. Once freely associated with the peroxisomal membrane, NBR1 molecules can interact through coiled-coil domains \cite{deosaran13, kirkin09}. While no specific NBR1-NBR1 interactions have been identified, membrane-bound proteins can form clusters through non-specific interactions.  The activity of the holin protein, leading to precise lysis timing \cite{ryan07} and showing cooperative effects across  distinct holin species \cite{wang03}, is thought to follow from clustering due to non-specific interactions. More generally, non-specific clustering of membrane-bound proteins can result from attractive lipid-mediated protein-protein interactions  \cite{heimburg96, gil98, lague01}. NBR1 oligomerization is also suggested by experiments  that show an important role for NBR1 in the formation of protein aggregates prior to their degradation by autophagy \cite{nicot14}.  Given a weak attractive interaction, the physical theory of phase separation  \cite{bray02} indicates that a sufficiently high concentration of NBR1 on a membrane will lead to a concentrated NBR1 phase -- i.e. cluster formation.  Such formation of homo-oligomeric clusters following  NBR1 membrane association through the J region is our cluster-driven selectivity hypothesis.

While p62 has no identified membrane binding domain, its PB1 region has a strong affinity ($K_d = 4-10$ nM \cite{wilson03}) which can lead to association with corresponding NBR1 PB1 regions and also to self-association into p62 filaments \cite{bienz14, lamark03}. While p62 has two binding faces on its PB1 region, NBR1 only has one \cite{lamark03} and so cannot form filaments.

We know that knockdown of p62 significantly affects pexophagy \cite{deosaran13}, but the mechanism is unknown.  Given the weak affinity of p62 to ubiquitin, it appears unlikely that p62 competition for NBR1 binding to ubiquitin is significant. While p62 binding to NBR1 and subsequent polymerization of p62 through PB1 domains would increase the number of LIR domains associated with an organelle, this in itself would be in proportion to the amount of NBR1 associated with the organelle  and would not affect NBR1 clustering.   However, polymer physics leads us to hypothesize that polymeric chains of p62 associated with NBR1 {\em reduce} NBR1 self-association through steric repulsion. Such a repulsion arises as the entropic contribution to the free-energy of polymers is decreased when brought close together \cite{hristova94}.  Such steric repulsion of membrane associated proteins can inhibit lipid phase separation \cite{scheve13}, or prevent growth of protein clusters \cite{sieber07}. 

Accordingly, we computationally model the association and disassociation of NBR1 on the surfaces of multiple peroxisomes -- as recruited by ubiquitin.  NBR1 can form clusters when its surface concentration is sufficiently high, and these clusters subsequently grow or shrink as determined by the available NBR1.  In the model, p62 can be recruited to membrane-associated NBR1, and subsequently polymerize. Because of steric repulsion, NBR1 that is associated with p62 does not participate in cluster formation or growth.

\subsection{Rates of NBR1 and p62 association}
While some equilibrium association constants are determined for NBR1 and p62, kinetic rate constants have not yet been measured. We use diffusion-limited association rates \cite{berg77, ghosh07}. In our model, we require rates for NBR1 to bind to ubiquitin targets on the surface of a peroxisome, or for p62 to bind to NBR1 targets on the surface of a peroxisome. The diffusion-limited arrival rate is known for arrival at small circular targets on a larger sphere \cite{berg77}, and within our model it is used for the arrival rate of NBR1 and p62. For NBR1, we use an association rate of NBR1 to each peroxisome
\begin{equation}
	\label{eq:jonnbr1}
	J_{on,NBR1} = 4\pi R \, \rho_{NBR1} D_{NBR1} \frac{ N_{ub} s_{ub}}{N_{ub}s_{ub} + \pi R},
\end{equation}
with $D_{NBR1}$ is the bulk NBR1 diffusivity, $\rho_{NBR1}$ is the bulk NBR1 concentration, $R$ the peroxisome radius, $N_{ub}$ the number of  ubiquitin on the peroxisome, and $s_{ub}$ the target ubiquitin radius. As discussed in the model motivation, we have sufficiently small bulk concentrations of NBR1 and p62 that a negligible fraction of ubiquitin are occupied, so the number of ubiquitin available for binding remains constant as NBR1 bind. We assume that NBR1 transiently bound to ubiquitin immediately associate with the peroxisomal membrane using their J regions. This allows significant NBR1 to accumulate on the peroxisome.

For p62, we similarly have diffusion-limited rates to surface associated NBR1
\begin{equation}
	\label{eq:jonp62}
	J_{on, p62}(\ell) = 4\pi R \rho_{p62} D_{p62}  \frac{ N_{NBR1}(\ell) s_{NBR1}}{N_{vapour}s_{NBR1}  + \pi R},
\end{equation}
where $D_{p62}$ is the bulk p62 diffusivity, $\rho_{p62}$ is the bulk p62 concentration, $s_{NBR1}$ is the (target) NBR1 radius, $N_{vapour}$ is the total number of NBR1 on the peroxisome surface that are not in clusters, and $N_{NBR1}(\ell)$ is the number of NBR1 with a p62-chain of length $\ell$.  Association extends the p62 chain length to $\ell+1$.

\subsection{NBR1 and p62 dissociation}
NBR1 on the surface of the peroxisome can dissociate from the membrane and return to the cytosol. We also model this as a diffusion-limited process, and for circular targets on a larger sphere, the dissociation rates have been determined \cite{ghosh07}. From Ghosh {\em et al} \cite{ghosh07}, the effective dissociation rate for NBR1 is 
\begin{equation}
	\label{eq:joffnbr1}
	J_{off,NBR1} = k_{off, NBR1} N_{vapour} ( 1- \gamma_{NBR1}),
\end{equation}
where $k_{off, NBR1}$ is the dissociation rate of NBR1 from the membrane and $\gamma_{NBR1} \equiv N_{ub} s_{ub}/(N_{ub} s_{ub} + \pi R)$ is the fraction of NBR1 that immediately rebind \cite{ghosh07}.  By equating the on and off rates of NBR1 to the peroxisome surface, $J_{on,NBR1}$ and $J_{off,NBR1}$, we see that the steady-state $N_{vapour}$ is proportional to $N_{ub}$, and otherwise independent of the peroxisomal radius $R$:
\begin{equation}
	\label{eq:steadystateNBR1}
	N_{steady-state \ vapour} = \frac{ 4 D_{NBR1} s_{ub} \rho_{NBR1}}{k_{off,NBR1}} N_{ub}.
\end{equation}
Equivalently, the steady-state surface concentration of NBR1 is proportional to that of ubiquitin.
 
In our model we assume that when NBR1 dissociates, any associated p62 chains dissociate as well. In addition, PB1 bonds (p62-p62 or p62-NBR1) within membrane-associated polymers will each break at a rate
\begin{equation}
	\label{eq:joffp62}
	J_{off,p62} = k_{off, p62} ( 1- \gamma_{p62}),
\end{equation}
where $\gamma_{p62} \equiv N_{vapour} s_{NBR1}/(N_{vapour} s_{NBR1} + \pi R)$ \cite{ghosh07}. When a PB1 bond breaks, the portion of the p62 chain beyond the bond (i.e. further from the NBR1 than the bond) dissociates. Since Eqn.~\ref{eq:joffp62} is the rate per PB1 bond, the rate of any PB1 in a chain of length $\ell$ breaking is proportional to $\ell$

\subsection{NBR1 cluster formation}
Large domains or clusters are generally only thermodynamically stable above some saturation density $\sigma_{\infty}$ of particles. Smaller clusters are less stable and require a higher density of particles to avoid shrinkage and evaporation; this is known as the Gibbs-Thomson effect \cite{krishnamachari96}. The Gibbs-Thomson effect is due to the increased curvature of the edge of a small cluster compared to a large cluster. Qualitatively, the increased curvature both reduces local bonding and allows particles more directions to escape. In equilibrium, these lead to a higher vapor concentration near the cluster edge to balance the increased escape rate. We need to consider the Gibbs-Thompson effect for small NBR1 clusters. 

Before cluster nucleation there will be freely diffusing NBR1 on a peroxisome of area $4 \pi R^2$ with surface concentration $\sigma$ and total NBR1 of $N= 4 \pi R^2 \sigma$, where $R$ is the peroxisomal radius. After cluster nucleation, there will be a cluster of radius $r$ with $N_{clust}$ NBR1 in equilibrium with a surface concentration $\sigma_{gt}$. Since the cluster is small, the surface concentration will satisfy $N_{surf} = 4 \pi R^2 \sigma_{gt}$. Since the number of NBR1 doesn't change, we have $N = N_{clust}+N_{surf}$. We also satisfy the Gibbs-Thomson effect \cite{krishnamachari96}, with the cluster radius $r$,
\begin{equation}
	\label{eq:formationsecond}
	\sigma_{gt} = \sigma_{\infty}\left(1 + \frac{\nu}{r}\right),
\end{equation}
where $\nu$ is a constant ``capillary length'' associated with NBR1 clusters. 

We cannot simply allow nucleation when $\sigma \geq \sigma_{gt}$, since the original surface concentration must also provide the NBR1 for the cluster formation. If $b$ is the area per NBR1 in a cluster, then the cluster area $\pi r^2 = b N_{clust}$. Since $\sigma = (N_{clust} + N_{surf})/(4 \pi R^2)= N_{clust}/(4 \pi R^2) + \sigma_{gt}$, then we require 
\begin{equation}
	\label{eq:intermediate}
	\sigma = \frac{r^2}{4R^2 b} + \sigma_{\infty}\left(1 + \frac{\nu}{r}\right).
\end{equation}

While we can therefore accommodate a range of possible cluster sizes $r$, there will be a smallest $r$ determined by minimizing  Eqn.~\ref{eq:intermediate} with respect to $r$. This determines a critical (minimal) surface concentration that allows for cluster nucleation, $\sigma^\ast$, where 
\begin{equation}
	\label{eq:minimumc}
	\sigma^\ast=\sigma_{\infty} + 3\left(\frac{\sigma_{\infty}\nu}{4\sqrt{b}R}\right)^{2/3}.
\end{equation}
We see that the minimum supersaturation required for nucleation is lower for larger peroxisomes \cite{krishnamachari96}. 
We also determine the cluster size after nucleation, $N_{clust}^\ast=\pi R^{4/3} \left({2\sigma_{\infty}\nu}/{\sqrt{b}}\right)^{2/3}$.

Following our assumption that NBR1 associated with p62 does not participate in cluster formation, only NBR1 with no associated p62 chain ($N_{NBR1}(l=0)$) contribute to the concentration required for nucleation in Eqn.~\ref{eq:minimumc}, so that we must have
\begin{equation}
\label{eq:thresholdconcentration}
	\sigma^\ast = N_{NBR1}(0)/(4 \pi R^2).
\end{equation}

We have assumed that each peroxisome will harbour either one or zero clusters. For other small biological systems with clusters, including bacterial holin domains \cite{ryan07, white11} and yeast polarity clusters \cite{howell12}, multiple clusters rapidly resolve to a single cluster. We also note that any supersaturation will be quickly absorbed by the first cluster to nucleate, suppressing further cluster nucleation by Eqn.~\ref{eq:minimumc}. 

\subsection{NBR1 cluster growth}
Existing NBR1 clusters can gain NBR1 and grow, or lose NBR1 and shrink. To determine the growth of an existing NBR1 cluster on a peroxisome we adapt the derivation in Appendix B of \cite{brown15}.

Only NBR1 without a p62 chain (with $\ell = 0$) can contribute to cluster growth. The number of such NBR1 on the peroxisome surface, $N(0)$, divided by the surface area $4\pi R^2$, gives a surface concentration $f_0$. In principle, $f_0$ is a spatial field over the peroxisome surface, varying depending on location. The dynamics of $f_0$ is then described by the partial differential equation
\begin{equation}
	\label{eq:cgderive1}
	\frac{df_0}{dt} = D_s\nabla^2 f_0 + \frac{J_{on,NBR1}}{4\pi R^2} - \frac{J_{off,NBR1}}{4\pi R^2} - \frac{J_{on,p62}(0)}{4\pi R^2} + 	\frac{1}{4\pi R^2}\sum_{\ell=1}^{\infty}J_{off,p62}N_{NBR1}(\ell). 
\end{equation}
The first term on the right hand side captures diffusion of NBR1 on the surface, with $D_s$ the surface diffusivity and $\nabla^2$ a two-dimensional Laplacian. The second term describes the arrival of NBR1 from the cytosol. The third term represents the dissociation of NBR1 from the surface. The fourth term gives the addition of p62 to an NBR1 with no p62, so that it has p62 and can no longer participate in cluster growth. The fifth term reflects the dissociation of p62 chains from NBR1, which allows those NBR1 to then participate in cluster formation.

After substituting the full expressions for $J_{on,NBR1}$ from Eqn.~\ref{eq:jonnbr1}, $J_{off,NBR1}$ from Eqn.~\ref{eq:jonp62}, $J_{off,NBR1}$ from Eqn.~\ref{eq:joffnbr1}, and $J_{off,p62}$ from Eqn.~\ref{eq:joffp62} into Eqn.~\ref{eq:cgderive1}, and assuming  close to steady state (i.e. $df_0/dt = 0$) we obtain 
\begin{equation}
	\label{eq:cgderive2}
	D_s\nabla^2 \tilde{f}_0 = a\tilde{f}_0,  
\end{equation}
where we define $\tilde{f}_0 \equiv f_0 - w$, 
\begin{equation}
	w = \frac{1}{a}\left[ \frac{\rho_{NBR1}D_{NBR1}N_{Ub}s_{Ub}}{R(N_{Ub}s_{Ub} + \pi R)} + \frac{k_{off,p62}}{4\pi R^2}\left(1 - 	\frac{N_{vapour}s_{NBR1}}{N_{vapour}s_{NBR1} + \pi R}\right)\sum_{\ell=1}^{\infty}N_{NBR1}(\ell)\right],
\end{equation}
and
\begin{equation}
	a = k_{off,NBR1}\left(1 - \frac{N_{Ub}s_{Ub}}{N_{Ub}s_{Ub} + \pi R}\right) + \frac{4 \pi R\rho_{p62}D_{p62}S_{NBR1}}		{N_{vapour}s_{NBR1} + \pi R}.
\end{equation}

Eqn.~\ref{eq:cgderive2} is solved in Appendix B of \cite{brown15} to determine the net flux to the cluster. This determines the dynamics of a cluster with $N_{clust}$ molecules on a  peroxisome of radius $R$,
\begin{equation}
	\label{eq:dynamical}
	\frac{dN_{clust}}{dt} = 4\pi a R^2\left[w - f_{\infty}\left(1 + \nu\sqrt{\frac{\pi}{bN_{clust}}}\right)\right].
\end{equation}
We use Eqn.~\ref{eq:dynamical} to determine the change in time of the size of every NBR1 cluster in our model.

\subsection{Kinetic model}
We implement our kinetic rates to continually update the NBR1 in our system.  Approximately 50$\%$ of NBR1 colocalizes with catalase and PMP70 \cite{deosaran13}, a peroxisome matrix and membrane protein, respectively, indicating that NBR1 dynamics on peroxisomes are probably not significantly buffered by other cellular processes. For every peroxisome, we track each $N_{NBR1}(\ell)$ and $N_{clust}$. The peroxisomal NBR1 is then $N_{peroxisomal} = N_{clust} + \sum_\ell N_{NBR1}(\ell)$, and we can sum that over all peroxisomes to obtain $N_{tot,peroxisomal}$. We conserve the total amount of NBR1 in the system, so that $N_{tot,bulk} = N_{tot} - N_{tot,peroxisomal}$ and the bulk density $\rho_{NBR1} = N_{tot,bulk}/V$, where $V$ is the total system volume. 

Only approximately 10$\%$ of p62 colocalizes with peroxisomes \cite{deosaran13}, which is consistent with the many roles of p62 for autophagy \cite{rogov14} as well as other cellular pathways \cite{komatsu12}. As a result, we expect that uptake by peroxisomes of p62 will not significantly change cytosolic concentrations. Accordingly, we hold p62 concentrations constant. 

\subsection{Selectivity calculation}
\label{sec:selectivity}
In Fig.~\ref{fig:figure2}(b), we show how the remaining peroxisomal volume fraction depends on size-selectivity when a fixed number fraction of peroxisomes are degraded.  With zero selectivity, with targets randomly chosen, the remaining volume and number fractions are proportional. With maximum selectivity, i.e., with only the larger peroxisomes selected for degradation, the volume fraction remaining at number fraction $r$ is 
\begin{equation}
\label{eq:vmax}
	v_{max} = (4/3)\pi \int_{R_{min}}^{R_{max}}R^3 P(R) dR/V_i,
\end{equation}
where $R_{min}$ is the radius of the smallest peroxisome, $R_{max}(r)$ is the radius of the largest peroxisome not selected for degradation, $P(R)$ is the distribution function of peroxisomal radii, and $V_i$ is the total initial peroxisomal volume. The corresponding number fraction is 
\begin{equation}
	\label{eq:pmax}
	r = \int_{R_{min}}^{R_{max}}P(R)dR,
\end{equation}
and we see that $R_{max}$ is determined by $r$. 

\subsection{Parameters, initial conditions, and numerical details}
The radius of a globular protein is approximately $R=0.066M^{1/3}$ for $R$ in nm with $M$ in Daltons \cite{erickson09}. We use this radius to estimate the size of diffusive targets on peroxisomes. For ubiquitin, of mass 8 kDa \cite{peng03, delft97}, $r_{ub}=1.32$ nm. For NBR1, with mass of approximately 107 kDa \cite{nicot14, waters09}, $r_{NBR1}=3.14$ nm.   

While the diffusivity of NBR1 or p62 have not been measured, we can scale the diffusivity of EYFP which is approximately $D_{YFP}=1$ $\mu$m$^2$/s \cite{kuhn11} (with mass $M_{YFP}=27$ kDa).  Assuming spherical (globular) proteins and corresponding Stokes-Einstein diffusivity, the diffusivity scales with inverse radius (or cube root of the mass), and we obtain $D_{NBR1}=0.63$ $\mu$m$^2$/s  and with a p62 mass of 62 kDa \cite{pankiv07, geetha02}  $D_{p62}=0.83$ $\mu$m$^2$/s.

Within our model, systems of many peroxisomes have peroxisome radii distributed exponentially, qualitatively like measured peroxisome size distributions \cite{liu11,vizeacoumar04}. In ensemble systems, $P(R)\sim e^{-R/R_s}$, where $P(R)$ is the probability of a peroxisome of radius $R$, and we use $R_s=0.1$ $\mu$m. The number of ubiquitin on a given peroxisome will be proportional to the surface area, $N_{ub}(R)=n_0(R/R_0)^2$, with the ubiquitin coefficient $n_0$ typically 100 unless otherwise stated, and $R_0=0.25$ $\mu$m.

We use a system volume $V=N_p v$, where $N_P$ is the number of peroxisomes, and $v=10$ $\mu$m$^3$ is the volume per peroxisome, unless otherwise stated. 300 peroxisomes has been reported as an average number for mammalian cells \cite{huybrechts09}. Therefore the volume inside a spherical cell of radius 10 $\mu$m, divided among 300 peroxisomes, is approximately 10 $\mu$m$^3$ per peroxisome.

For cluster formation, we assume the capillary length $\nu$ is the size of a single NBR1 protein, so $\nu = r_{NBR1} = 3.14$ nm, and that the area per molecule is $b=\nu^2=9.86$ nm$^2$. This is consistent with capillary lengths of one \cite{krishnamachari96} and several \cite{strobel01} particle widths for 2$d$ and 3$d$ systems, respectively. The vapour pressure $\sigma_{\infty}$ is taken to be 10 $\mu$m$^{-2}$ on the peroxisome membrane. For a typical peroxisome of radius $R=0.25$ $\mu$m, this is approximately a single molecule on the peroxisomal surface.

We use $k_{off,NBR1}=0.1$ $\mu$m s$^{-1}$ for the dissociation rate of NBR1 from the peroxisome membrane, which yields a $K_d$ value similar to a 20 nM value measured for amphipathic helices \cite{stahelin03}. We choose $k_{off,p62} = 0.04$ s$^{-1}$ for the dissociation rate of p62 from NBR1 and p62 (interaction through the PB1 domain \cite{wilson03,lamark03,ciuffa15}) using the $K_d$ range of 4 - 10 nM for PB1-PB1 bonds \cite{wilson03}.

\subsection{Model limitations}
In this section we address some of the limitations of our approach. Our results should not be qualitatively affected by these limitations, whereas precise quantitative predictions would need a more realistic and dynamic cellular geometry, precise parameterization, and a fully stochastic multi-scale approach.

We have used deterministic dynamics in our modeling approach; this was necessary for computational efficiency since our system spans many length and time scales. Our model does not include stochastic effects for the change in molecule number on peroxisomes or in clusters, or for the nucleation of clusters. Change in molecule numbers on peroxisomes and in clusters is determined by the net flux, with frequent molecular association and dissociation. Our deterministic approach reflects an average behavior, and has two limitations. First it does not account for the discrete nature of molecules on peroxisomes or in clusters. The discreteness will affect our nucleation threshold, and we could impose that the number of molecules in a nucleated cluster, $N^\ast_{clust}$, is at least one. Since $N^\ast_{clust}$ is strongly $R$ dependent, a larger minimum cluster size due to discreteness will raise the minimum peroxisome size $R^\ast$ that nucleates clusters. This will enhance our predicted size selectivity effect, and so amounts to a conservative approximation. Second, we assume that cluster nucleation occurs deterministically at the threshold.  Since nucleation rates typically strongly increase with concentration, the threshold approximation is expected to reproduce the qualitative nucleation behavior. 

We have also abstracted the cellular context into uniform concentrations of bulk solutes, rather than an exact stochastic particle-based approach that includes cellular synthesis and degradation --- again for computational efficiency. A significant additional advantage of a uniform solute approximation is that it does not require us to model the precise cellular geometry, such as peroxisome locations. This does mean, however, that we cannot treat screening effects between peroxisomes. Identical peroxisomes will behave identically in our model but not within the cell. 

An important stochastic effect that we do not include is the downstream autophagy process that removes peroxisomes from the system. We would expect this (missing) process to limit the growth of NBR1 clusters.  Without it, NBR1 clusters, according to Eq.~\ref{eq:dynamical}, could grow indefinitely. For the parameters of our model, clusters on typical peroxisomes of radius $R = 0.25$ $\mu$m could reach 10--20$\%$ surface coverage at late times. The absence of downstream degradation is most significant for the maximum selectivity case of Fig.~\ref{fig:figure2}, since in that case the largest remaining peroxisome should retain its cluster until degradation. 

We have simplified our receptor dynamics as much as possible. For example, no p62-associated NBR1 will form clusters, but no NBR1 in clusters will associate with p62. We have also assumed that NBR1 is recruited to peroxisomes only by membrane-associated ubiquitin. It is also possible that NBR1 could also be recruited by already membrane-associated ubiquitin. Such direct recruitment could affect the cluster size dynamics \cite{Gov06}.

As noted previously, many of our parameters are generic estimations, as they have not been measured directly. While the qualitative physics of nucleation and diffusion will be unchanged by large parameter changes, the degree of  size-selectivity and the timing and extent of cluster nucleation is parameter dependent. We expect that by not including stochastic effects, we will have effectively shifted the appropriate parameter values. Since our parameters are rough estimations in any case, this does not change our qualitative results and conclusions.

\section*{Funding}
ADR  thanks the Natural Science and Engineering Research Council of Canada (NSERC) for operating grant support (RGPIN-2014-06245). We thank ACENET for computational resources. AIB  also thanks NSERC, the Killam Trusts, and the Walter C. Sumner Memorial Foundation for fellowship support. 

\section*{Acknowledgments}
We thank Peter Kim for helpful discussions. 

\bibliography{references}

\end{document}